\documentclass[twocolumn,showpacs,preprintnumbers,amsmath,amssymb]{revtex4}

\usepackage{epsfig}
\usepackage{subfigure}
\usepackage{bm}

\usepackage[normalem]{ulem} 
\usepackage[dvips]{color} 
\renewcommand\sout{\bgroup \color{red} \ULdepth=-.5ex \ULset}

\newcommand{\be}{\begin{equation}}
\newcommand{\ee}{\end{equation}}
\newcommand{\bea}{\begin{eqnarray}}
\newcommand{\eea}{\end{eqnarray}}

\newcommand{\nn}{\nonumber\\}

\begin{document}

\title{Dilepton production in nucleus-nucleus collisions at top SPS energy within
the Parton-Hadron-String-Dynamics (PHSD) transport approach}

\author{O.~Linnyk}

\email{Olena.Linnyk@theo.physik.uni-giessen.de}

\affiliation{%
 Institut f\"ur Theoretische Physik, %
  Universit\"at Giessen, %
  35392 Giessen, %
  Germany %
}

\author{E.~L.~Bratkovskaya}%
\affiliation{%
 Institut f\"ur Theoretische Physik, %
 Johann Wolfgang Goethe University, %
 60438 Frankfurt am Main, %
 Germany; %
Frankfurt Institute for Advanced Studies, %
 60438 Frankfurt am Main, %
 Germany; %
 }

\author{V.~Ozvenchuk}%
\affiliation{%
Frankfurt Institute for Advanced Studies, %
 60438 Frankfurt am Main, %
 Germany %
}

\author{W.~Cassing}
\affiliation{%
 Institut f\"ur Theoretische Physik, %
  Universit{\"a}t Giessen, %
  35392 Giessen, %
  Germany %
}%

\author{C. M.~Ko}%
\affiliation{%
Cyclotron Institute and Department of Physics and Astronomy, %
Texas A\&M University, %
College Station, TX
77843-3366, USA}

\date{\today}

\begin{abstract}
Dilepton production in In+In collisions at 158 A$\cdot$GeV is
studied within the microscopic parton-hadron-string dynamics (PHSD)
transport approach that incorporates explicit partonic
degrees-of-freedom, dynamical hadronization as well as the more
familiar hadronic dynamics in the final reaction stages. A
comparison to the data of the NA60 Collaboration shows that the
{measured} dilepton yield is well described by including the
collisional broadening of vector mesons, while simultaneously
accounting for the electromagnetic radiation of the strongly coupled
quark-gluon plasma (sQGP) via off-shell quark-antiquark
annihilation, quark annihilation with additional gluon
Bremsstrahlung and the gluon-Compton scattering mechanisms. In
particular, the spectra in the intermediate mass range (1~GeV~$\le
M\le2.5$ GeV) are dominated by quark-antiquark annihilation in the
nonperturbative QGP. Also, the observed softening of the transverse
mass spectra at intermediate masses (1~GeV~$\le M\le2.5$~GeV) is
approximately reproduced. Furthermore, for dileptons of low masses
($M<0.6$~GeV), we find a sizable contribution {from} the quark
annihilation with additional gluon bremsstrahlung, thus providing
another possible window for probing the properties of the sQGP.
\end{abstract}

\pacs{%
25.75.-q, 13.60.Le, 12.38.Mh, 14.40.Lbp, 14.65.Dw %
}

\keywords{%
Relativistic heavy-ion collisions\sep Meson production\sep
Quark-gluon plasma }

\maketitle

\section{Introduction}

Electromagnetic probes, i.e. dileptons and photons, are powerful
tools to explore the early hot, dense stage of heavy-ion collisions
as they are essentially unaffected by final{-}state interactions.
Through their invariant mass and momentum distributions, they carry
to the detector information about the conditions and properties of
the environment in which they are emitted, thus providing a glimpse
deep into the bulk of the strongly interacting matter created in
these collisions~\cite{Tserruya:2009zt,Fries:2002kt}. In particular,
dileptons have been suggested as probes of the quark-gluon plasma
(QGP) that is expected to be produced during the early stage of
heavy-ion collisions at Super-Proton-Synchrotron (SPS)
energies~\cite{Shuryak:1977ut,Shuryak:1978ij,Feinberg:1970tg,Feinberg:1976ua,Bjorken:1975dk}.

Recently, the NA60 Collaboration~\cite{NA60} has measured dileptons
from In+In collisions at $158$ A$\cdot$GeV and found that the
inverse slope parameter or effective temperature of the transverse
mass spectrum of dileptons in the intermediate mass region is lower
than that of dileptons at lower masses, which are dominantly of
hadronic origin. This might be explained if the dilepton spectrum at
invariant masses above 1~GeV is essentially due to partonic channels
in the QGP~\cite{Rapp:2009yu,Linnyk:2009nx,Linnyk:2010ub}. In this
case, the softening of the transverse mass spectrum with increasing
invariant mass implies that the partonic channels occur dominantly
before the collective radial flow has developed.

Since dileptons are emitted over the entire history of the heavy-ion
collision, -- from the initial nucleon-nucleon collisions through
the hot and dense phase and to the hadron decays after freeze-out,
-- microscopic covariant transport models {are very useful for}
disentangl{ing} the various sources that contribute to the final
dilepton spectra seen {in experiments}. The assumption that the
dilepton spectrum at masses above 1~GeV might be dominated by
radiations from the QGP was supported by studies within the
Hadron-String-Dynamics (HSD) transport approach~\cite{Cass99}, which
has shown~\cite{Bratkovskaya:2008bf} that the measured dilepton
yield at low masses ($M\le1$~GeV) can be well explained by dilepton
production from hadronic interactions and decays, while there is a
discrepancy between the HSD results and the data in the mass region
above 1~GeV. The {excess} seen for $M>1$~GeV could not be accounted
for by hadronic sources in HSD with or without medium effects and
might be interpreted as a signal for the existence of partonic
matter already in heavy-ion collisions at 158~A$\cdot$GeV incident
energy. Indeed, results from {model studies by} Dusling and Zahed
\cite{Dusling} as well as Renk and Ruppert \cite{Renk:2008prc} have
indicated that this excess could be due to partonic channels, i.e.
primarily to $q \bar{q}$ annihilation. On the other hand, this
dilepton excess has been attributed by Rapp and collaborators to
multi-meson production channels (denoted shortly as $4
\pi$-contribution \cite{RH:2008lp}). These different interpretations
{of the experimental data} are still being extensively debated. In
this respect{,} the physics of dilepton transverse momentum spectra
can be especially relevant~\cite{Song:2010fk}. Due to the
nonequilibrium nature of heavy-ion {collisions}, a clarification
within a transport approach that incorporates dilepton production
from the (non-equillibrium) partonic phase, hadronic decays and the
microscopic secondary hadronic interactions -- including the ``$4
\pi$" channels -- {thus} appears appropriate.

Another open question to be answered by the microscopic transport
calculations is the existence of other `windows' in the phase space
for observing dileptons from the {quark-gluon} plasma over the
hadronic sources. It has been originally suggested that the
substantial thermal yield from the deconfined phase existed in the
invariant mass region between the $\phi$ and $J/\Psi$
peaks~\cite{Shuryak:1978ij}, while the spectrum at lower masses was
dominated by meson decays. On the other hand, the
calculations~\cite{Gallmeister:1999dj,Gallmeister:2000ra} of the
thermal {dilepton} yield from $q\bar q$ {annihilation} in a blast
wave model in comparison to {that from} a hadronic cocktail, {the}
Drell-Yan {mechanism} and the correlated semileptonic decays of open
charm have found a possible second region of the phase space for the
observation of {this} thermal source at masses $\approx 0.3-0.6$~GeV
and low transverse momentum.

The Parton-Hadron-String Dynamics~\cite{CasBrat,BrCa11} (PHSD)
transport approach, which incorporates the relevant off-shell
dynamics of vector mesons and the explicit partonic phase in the
early hot and dense reaction region as well as the dynamics of
hadronization, allows for a microscopic study of various dilepton
{production} channels {in nonequilibrium matter}. The PHSD off-shell
transport approach is particularly suitable for this investigation,
since it incorporates various scenarios for the modification of
vector mesons in a hot and dense medium, seen experimentally in the
enhanced production of lepton pairs in the invariant mass range $0.3
\leq M \leq 0.7$ GeV/$c^2$. In the present work, we calculate
dilepton production from the partonic and hadronic sources within
PHSD {by including} the multi-meson channels and the partonic
channels {besides the usual hadron decay channels}. By consistently
treating in the same microscopic transport framework both partonic
and hadronic phases of the collision system, we are aiming to
determine the relative importance of different dilepton production
mechanisms and to point out the regions in phase space where
partonic channels are dominant.

The paper is organized as follows. In Sec. II, we give a brief
description of the PHSD approach. We then describe in Sec. III the
partonic sources of dilepton production incorporated in PHSD and in
Sec. IV dilepton production by (in-medium) hadrons and in
multi-meson processes. In Sec. V, we compare the results of the
calculations to the available experimental data. Finally, the
conclusions are given in Sec. VI.


\section{PHSD transport approach}
\label{section.PHSD}

To address dilepton production in a hot and dense medium as created
in heavy-ion collisions, we employ an up-to-date relativistic
transport model, i.e. the Parton Hadron String
Dynamics~\cite{CasBrat,BrCa11} (PHSD). PHSD consistently describes
the full evolution of a relativistic heavy-ion collision from the
initial hard scatterings and string formation through the dynamical
deconfinement phase transition to the quark-gluon plasma as well as
hadronization and to the subsequent interactions in the hadronic
phase.

In the hadronic sector, PHSD is equivalent to the
Hadron-String-Dynamics (HSD)  transport approach
\cite{Cass99,Brat97,Ehehalt} that has been used for the description
of $pA$ and $AA$ collisions from SIS to RHIC energies and has lead
to a fair reproduction of {measured} hadron abundances, rapidity
distributions and transverse momentum spectra. In particular, {the}
HSD incorporates off-shell dynamics for vector
mesons~\cite{Cass_off1} and a set of vector-meson spectral
functions~\cite{Brat08} that covers possible scenarios for their
in-medium modification.

The transition from {the} partonic to hadronic degrees of freedom is
described by covariant transition rates for the fusion of
quark-antiquark pairs to mesonic resonances or three quarks
(antiquarks) to baryonic states, i.e. dynamical hadronization. Note
that due to the off-shell nature of partons on one hand and the
resulting hadrons on the other, the hadronization process obeys all
conservation laws (i.e. 4-momentum conservation, flavor current
conservation) in each event, the detailed balance relations, and the
increase in {the} total entropy $S$. The transport theoretical
description of quarks and gluons in {the} PHSD is based on a
dynamical quasiparticle model (DQPM) for partons matched to
reproduce lattice QCD (lQCD) results {for a quark-gluon plasma} in
thermodynamic equilibrium. {The} DQPM provides {the} mean-fields for
gluons/quarks and {their} effective 2-body interactions for the
implementation to PHSD.

\begin{figure*}
\centering \subfigure[]{ \label{Quark3D}
\resizebox{0.51\textwidth}{!}{%
 \includegraphics{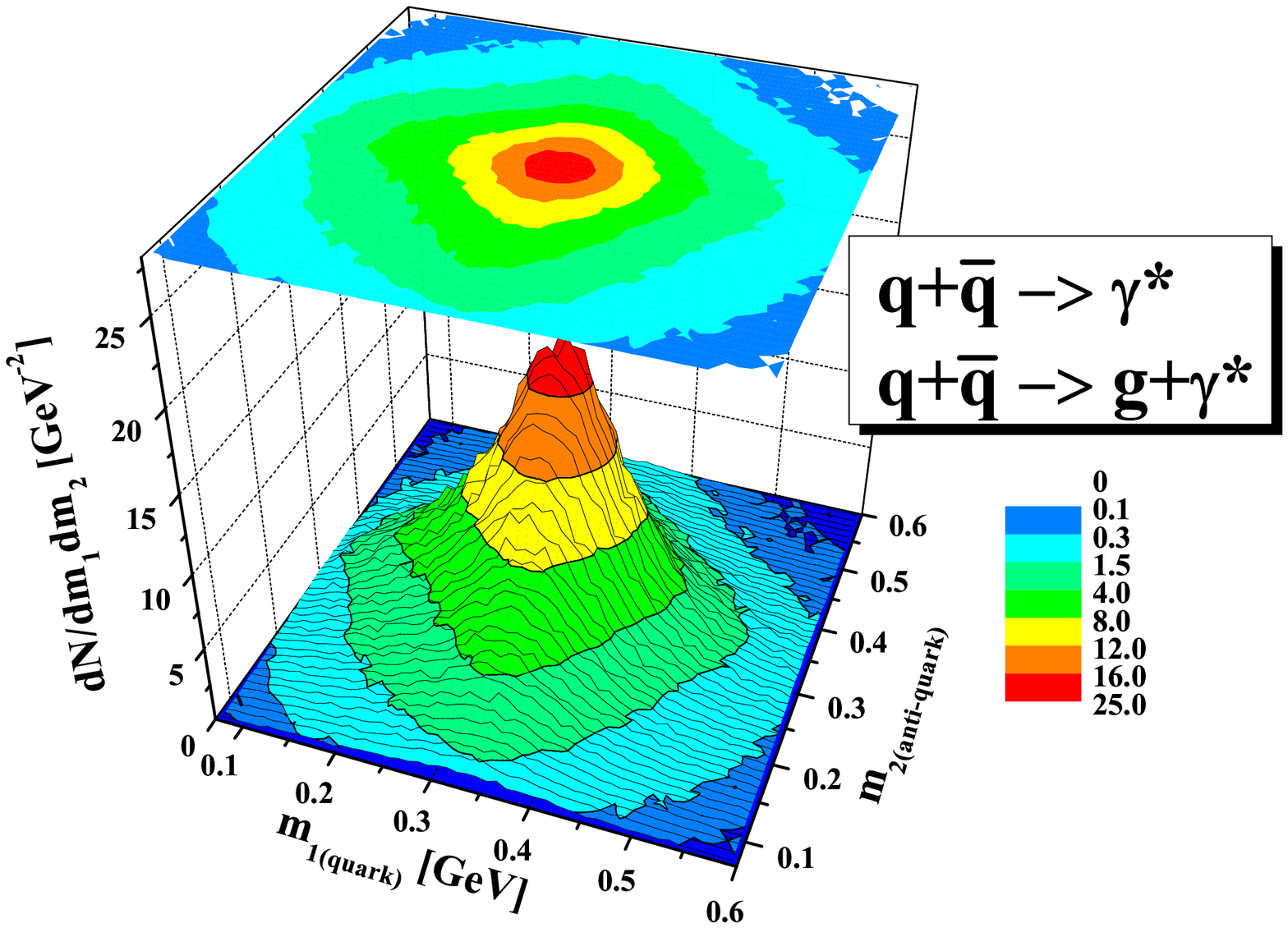}
} } \subfigure[]{ \label{Gluon3D}
\resizebox{0.46\textwidth}{!}{%
 \includegraphics{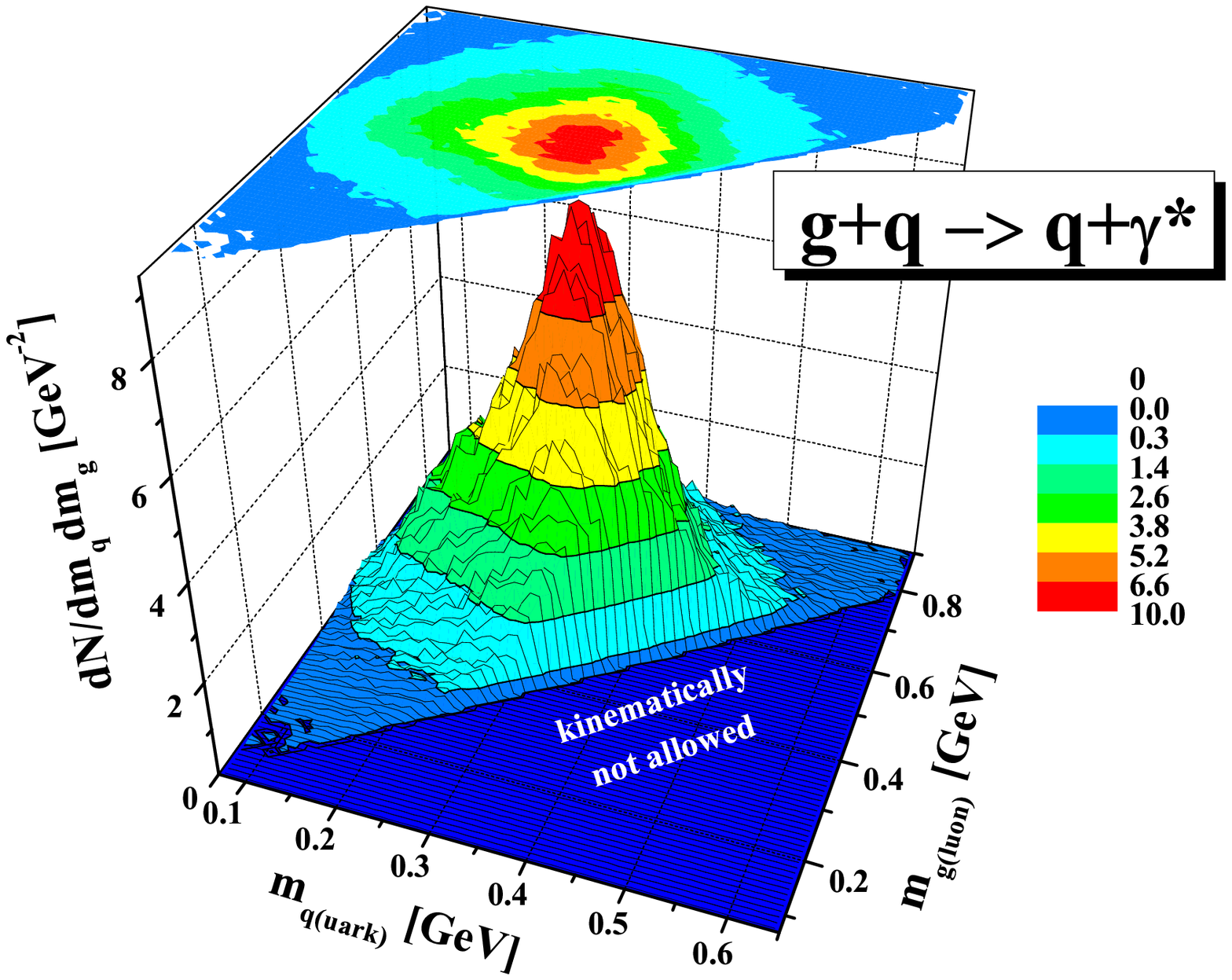}
} } \caption{(color online)  The number of  $q + \bar q$ (a) and
$q+g$ (b) collisions, in which a dilepton pair is produced, in a
central $In+In$ reaction at an incident energy of 158~A$\cdot$GeV
versus the masses of the (quasiparticle) quark and antiquark as
resulting from PHSD. }
\end{figure*}

We briefly recall the basic assumptions of the DQPM model (for
details about the DQPM model and the off-shell transport we refer to
Ref.~\cite{Cassing:2008nn}). Following Ref.~\cite{Andre05}{,} the
dynamical quasiparticle mass (for gluons and quarks) is assumed to
be given by the thermal mass in the asymptotic high-momentum regime,
which is proportional to the temperature $T$ {and} a running
coupling {$g(T/T_c)$} (squared), for which the following
parametrization is used
\begin{eqnarray}
 g^2(T/T_c) = \frac{48\pi^2}{(11N_c - 2 N_f)
\ln[\lambda^2(T/T_c-T_s/T_c)^2]}\ .
 \label{eq:g2}
\end{eqnarray}
Here $N_c = 3$ stands for the number of colors while $N_f$ denotes
the number of flavors. The parameters controlling the infrared
enhancement of the coupling $\lambda $ and $T_s $ have been fitted
in Ref.~\cite{BrCa11} to recent lQCD results for the entropy density
$s(T)$. An almost perfect reproduction of the energy density
$\varepsilon(T)$ and the pressure $P(T)$ from lQCD is achieved as
well (cf.~\cite{BrCa11}).

In line with Ref.~\cite{Andre05}, the parton spectral functions are
no longer $\delta-$functions in the invariant mass squared but taken
as
\begin{eqnarray}
 \rho_j(\omega)
 =
 \frac{\gamma_j}{ E_j}
 \left(
   \frac{1}{(\omega-E_j)^2+\gamma_j^2} - \frac{1}{(\omega+E_j)^2+\gamma_j^2}
 \right)
 \label{eq:rho}
\end{eqnarray} separately for quarks and gluons ($j=q,\bar{q},g$).
With the convention $E^2_j(p) = \bm p^2+M_j^2-\gamma_j^2$, the
parameters $M_j^2$ and $\gamma_j$ are directly related to the real
and imaginary parts of the  retarded self-energy, e.g. $\Pi_j =
M_j^2-2i\gamma_j\omega$.

The width  for gluons and quarks (for vanishing chemical potential
$\mu_q$) is adopted in the form
\begin{eqnarray} \label{eq:gamma}
  \gamma_g(T)
  =
  \frac{3 g^2 T}{8 \pi} \,  \ln\left( \frac{2c}{g^2}\right)  \, , \
    \gamma_q(T)
  =
   \frac{g^2 T}{6 \pi} \,  \ln
  \left(\frac{2c}{g^2}\right)
  \, ,
\end{eqnarray}
where $c=14.4$ (from Ref.~\cite{Peshier:2005pp}) is related to a
magnetic cut-off.

We stress that a non-vanishing width $\gamma$ is the main difference
{between} the DQPM {and} conventional quasiparticle
models~\cite{qp1}. Its influence is essentially seen in correlation
functions, e.g., in the stationary limit of the correlation in the
off-diagonal elements of the energy-momentum tensor $T^{kl}$ which
defines the shear viscosity $\eta$ of the
medium~\cite{Peshier:2005pp}. Here a sizeable width is mandatory to
obtain a small ratio of the shear viscosity to entropy density
$\eta/s$, which results in a roughly hydrodynamical evolution of the
partonic system in PHSD \cite{Cass08}. The finite width leads to
two-particle correlations, which are taken into account by means of
the {\em generalized}, off-shell transport
equations~\cite{Cass_off1} {that go} beyond the mean field or
Boltzmann approximation~\cite{Cassing:2008nn,Linnyk:2011ee}.

{The off-shell effect can be seen for example in Fig.~\ref{Quark3D}}
where the number of the $q+\bar q$ {and $q + g$} collisions -- in
which a dilepton pair can be formed -- is shown as a function of the
participating parton masses. The plots have been generated by a
simulation in PHSD for a central $In+In$ reaction at an incident
energy of 158~A$\cdot$GeV. The maximum of the distribution indicates
the average pole mass of the quark/gluon, while the width correlates
with the average width of their spectral function. The values for
the masses and widths are in agreement with those from the DQPM fit
to the lattice data for the temperatures in the range $\approx 1-2\
T_c$.

\begin{figure}[hbt]
\centering
\resizebox{0.49\textwidth}{!}{%
 \includegraphics{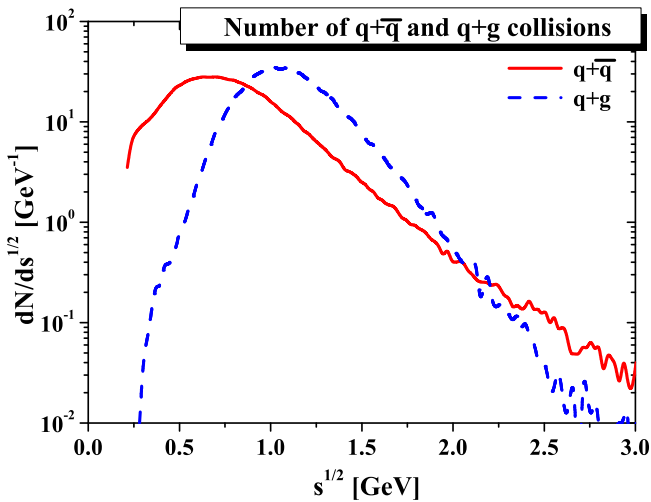}
} \caption{(color online) Number of parton collisions per event in a
central $In+In$ reaction at an incident energy 158~A$\cdot$GeV
versus the invariant energy $\sqrt{s}$ of the elementary partonic
collision as simulated in PHSD. The number of $q+\bar q$ collisions
is given by the solid (red) line while that of $q+g$ collisions is
given by the dashed (blue) line.} \label{dNdS}
\end{figure}

\begin{figure*}
\includegraphics[width=\textwidth]{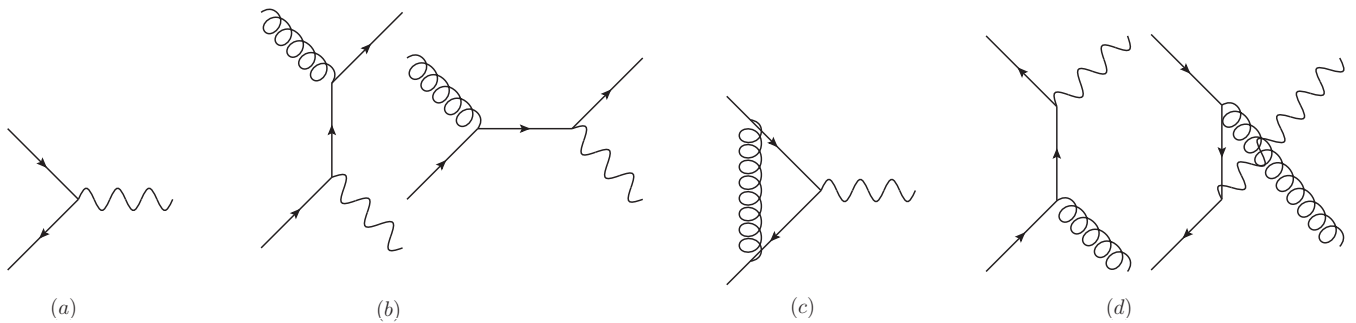}
\caption{Diagrams contributing to dilepton production from the
 QGP: (a) Drell-Yan mechanism, (b) gluon-Compton
scattering (GCS), (c) vertex correction, (d) gluon Bremsstrahlung
(NLODY), where virtual photons (wavy lines) split into lepton pairs,
spiral lines denote gluons, and arrows denote quarks. In each
diagram the time runs from left to right.} \label{diagrams}
\end{figure*}

\begin{figure*}
\begin{center}
\resizebox{\textwidth}{!}{%
 \includegraphics{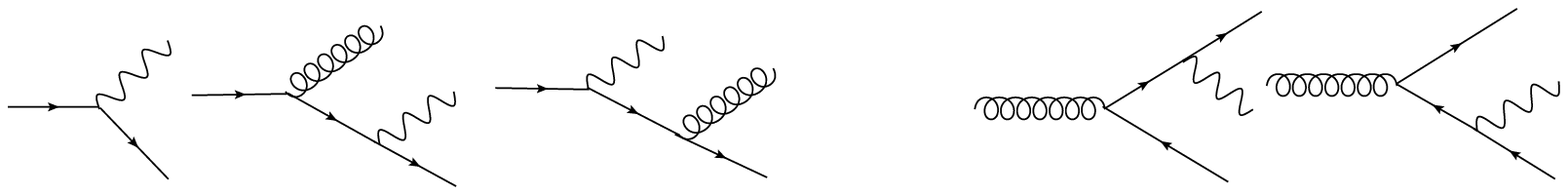}
} \caption{Diagrams contributing to dilepton production by virtual
quasi-particles in addition to those presented in
Fig.~\protect{\ref{diagrams}}. {Left}: the decay of a virtual quark;
 {Right}: the decay of a virtual gluon. Virtual
photons (wavy lines) split into lepton pairs, spiral lines denote
gluons, and arrows denote quarks.} \label{diagrams2}
\end{center}
\end{figure*}

For an illustration of the quark and gluon interactions in a
heavy-ion collision as generated in PHSD, we show in Fig.~\ref{dNdS}
the number of  {$q+{\bar q}$ (solid line) and $q+g$ (dashed line)}
collisions that can create dilepton pairs per event in a central
$In+In$ reaction at an incident energy of 158~A$\cdot$GeV versus the
invariant energy $\sqrt{s}$ of the elementary partonic collision.
One can see that the tails of the collision distributions calculated
in the PHSD transport are almost exponential, thus close to thermal.
On the other hand, the collisions at very low $\sqrt{s}$ are
suppressed. This `threshold effect' is due to the finite masses of
the dynamical quarks, antiquarks and gluons. Additionally, one
notices that the threshold is not sharp because of the rather broad
spectral functions (and therefore broad mass distributions) of the
colliding partons.


\section{Partonic sources of dileptons in PHSD}
\label{section.partonic}

In the scope of the one- and two-particle interactions, dilepton
radiation by the constituents of the strongly interacting QGP
proceeds via the elementary processes illustrated in
Figs.~\ref{diagrams} and~\ref{diagrams2}: the basic Born $q+\bar q$
annihilation mechanism, gluon Compton scattering ($q+g\to
\gamma^*+q$ and $\bar q+g\to \gamma^*+\bar q$), and quark +
anti-quark annihilation with gluon Bremsstrahlung in the final state
($q+\bar q\to g+\gamma^*$), virtual quark decay ($q \to q+
g+\gamma^*$) and virtual gluon decay ($g\to q+\bar q+\gamma^*$) . In
the on-shell approximation, one uses perturbative QCD cross sections
for the processes listed above. However, in the strongly interacting
QGP the gluon and quark propagators differ significantly from the
non-interacting propagators. Accordingly, the cross sections for
dilepton production in the partonic channels have been calculated in
Ref.~\cite{olena2010} in the DQPM model that had been fitted to
lattice QCD results in thermal equilibrium
before~\cite{Peshier:2005pp}.

The importance of finite mass corrections to the perturbative cross
sections has been stressed in Ref.~\cite{olena2010}. It was shown
that the finite quark and gluon masses can modify the magnitude as
well as the $M-$ and $p_T-$dependence of the cross sections of the
processes in Fig.~1 compared to the perturbative results for
massless partons (cf. Figs.~3 and 4 of Ref.~\cite{olena2010}). The
modifications are {large} at lower $M^2$ and at the edges of the
phase space. It was shown that the most prominent effect of the
quark masses on the dimuon production cross sections in the Born
mechanism ($q+\bar q\to \gamma^*$) was a sharp threshold value for
the invariant mass of the dilepton pair $M_{min}=m_1+m_2$. On the
other hand, the finite masses of the quark and antiquark produce
additional higher-twist corrections to the cross section, which
decrease with increasing $M^2$, so that the off-shell cross sections
approach the leading twist -- on-shell -- result in the limit of
high dilepton masses. In Fig.~4 of Ref.~\cite{olena2010}, an
analogous comparison for the $2\to2$ process $q+\bar q\to
\gamma^*+g$  was shown by plotting the off-shell (i.e. with finite
masses for the quarks and gluons) cross section for the quark
annihilation with gluon bremsstrahlung in the {final} state at
various values {of} the quark and gluon off-shellnesses (masses) and
the corresponding on-shell result. As found in
Ref.~\cite{olena2010}, the maximum pair mass shifts to a lower value
(in order to produce a massive gluon in the final state). For the
rest of the $M$ values, the effect of the quark and gluon masses is
about 50\%. For $m_{q/g}\to0$, the cross section approaches the
leading twist pQCD result.

The question of the effect of a finite parton width -- which
parametrizes the effect of their interaction rate and correlation,
including multiple scattering -- on dilepton rates in heavy-ion
collisions was addressed in Ref.~\cite{olena2010} by convoluting the
off-shell cross sections with phenomenological spectral functions
$A(m_q)$ and $A(m_g)$ for the quarks and gluons in the quark-gluon
plasma and with parton distributions in a heavy-ion collision
similar to those of Fig.~\ref{dNdS} {in} the {present} paper. The
finite width of the quasiparticles was found to have a sizable
effect on the dilepton production rates. In particular, the
threshold of the Drell-Yan contribution was "washed out". Also, the
shape and magnitude of the $2\to2$ processes ($q+\bar q\to
g+\gamma^*$ and $q+g\to q+\gamma^*$) {were} modified. One further
observed that the contribution of the gluon Compton process $q+g\to
q+\gamma^*$ to the rates was small compared to that of $q+\bar q$
annihilations.

In the {present} work, we implement the cross sections obtained in
Ref.~\cite{olena2010} into the PHSD transport approach in the
following way: Whenever the quark-antiquark, quark-gluon and
antiquark-gluon collisions occur in the course of the Monte-Carlo
simulation of the partonic phase in PHSD, a dilepton pair can be
produced according to the off-shell cross sections~\cite{olena2010},
which depend, in addition to the virtualities of the partons
involved, on the energy density in the local cell, in which the
collision takes place. The local energy density governs the widths
of the quark and gluon spectral functions as well as the strong
coupling (cf. Eqs.~(\ref{eq:g2}) and (\ref{eq:gamma}) that depend on
temperature $T$ which in turn is uniquely related to the energy
density by the lattice QCD equation of state). Numerically, one
finds from a PHSD simulation of a heavy-ion collision at SPS
energies that the running coupling $\alpha_S$ in the partonic phase
is often of order $O(1)$ and thus the contribution of the
higher-order Bremsstrahlung diagram is compatible in magnitude to
the Born term.


\section{Hadronic sources of dileptons in PHSD}
\label{section.hadronic}

In the hadronic sector, PHSD is equivalent to the
Hadron-String-Dynamics (HSD)  transport approach
\cite{Cass99,Brat97,Ehehalt}. The implementation of the hadronic
decays into dileptons ($\pi$-, $\eta$-, $\eta '$-, $\omega$-,
$\Delta$-, $a_1$-Dalitz, $\rho\to l^+l^-$, $\omega\to l^+l^-$,
$\phi\to l^+l^-$) in HSD (and PHSD) is described in detail in
Refs.~\cite{Brat08,Bratkovskaya:2008bf}. For the treatment of the
leptonic decays of open charm mesons and charmonia we refer to
Refs.~\cite{Manninen:2010yf,Linnyk:2008hp}. In the present paper, we
extend the hadronic sources {for} dilepton production to include
secondary multi-meson interaction{s} by incorporating the channels
$\pi \omega \to l^+l^-$, $\pi a_1\to l^+l^-$, {and} $\rho \rho \to
l^+l^-$.

The dilepton production by a (baryonic or mesonic) resonance $R$
decay in HSD and PHSD can be schematically presented in the
following way:
\begin{eqnarray}
 BB &\to&R X   \label{chBBR} \\
 mB &\to&R X \label{chmBR} \\
      && R \to  e^+e^- X, \label{chRd} \\
      && R \to  m X, \ m\to e^+e^- X, \label{chRMd} \\
      && R \to  R^\prime X, \ R^\prime \to e^+e^- X, \label{chRprd}
\end{eqnarray}
i.e. in a first step a resonance $R$ might be produced in
baryon-baryon ($BB$) or meson-baryon ($mB$) collisions
(\ref{chBBR}), (\ref{chmBR}). Then this resonance can couple to
dileptons directly (\ref{chRd}) (e.g., Dalitz decay of the $\Delta$
resonance: $\Delta \to e^+e^-N$) or decays to a meson $m$ (+ baryon)
or in (\ref{chRMd})  produce dileptons via direct decays ($\rho,
\omega$) or Dalitz decays ($\pi^0, \eta, \omega$). The resonance $R$
might also decay into another resonance $R^\prime$  (\ref{chRprd})
which later produces dileptons via Dalitz decay.  Note, that in the
combined model the final particles -- which couple to dileptons --
can be produced also via non-resonant mechanisms, i.e. 'background'
channels at low and intermediate energies or string decay at high
energies.

\subsection{In-medium modification of vector mesons}

While the  properties of hadrons are rather well known in free space
(embedded in the nonperturbative QCD vacuum), the masses and
lifetimes of hadrons in a baryonic and/or mesonic environment are
subject of current research that aims at achieving a better
understanding of the strong interaction and the nature of
confinement. For example, a broadening of the vector mesons can be
understood as a shortening of the lifetime of the vector mesons
$\rho$, $\omega$ and $\phi$ in the medium. In this context the
modification of hadron properties in nuclear matter are of
fundamental interest
(cf.~\cite{BrownRho,H&L92,Asakawa93,Shakin94,Klingl96}), since QCD
sum rules~\cite{H&L92,Asakawa93,Leupold} as well as QCD inspired
effective Lagrangian models~\cite{BrownRho,Shakin94,Rapp96,Peters}
predict significant changes, e.g., in the properties of the vector
mesons ($\rho$, $\omega$ and $\phi$) with the nuclear density
$\rho_N$ and/or temperature
$T$~\cite{Cass99,rapp5,Song:1994sj,Song:1995af}.

A  modification of the properties of vector mesons in the nuclear
medium was first seen experimentally in the enhanced production of
lepton pairs above known sources in nucleus-nucleus collisions at
SPS energies \cite{CERES,HELIOS}. As proposed in
Refs.~\cite{Li,Li96}, the observed enhancement in the invariant mass
range $0.3 \leq M \leq 0.7$ GeV/$c^2$ might be due to a shift of the
$\rho$-meson mass following the Brown/Rho scaling~\cite{BrownRho} or
the Hatsuda and Lee sum rule prediction~\cite{H&L92}.  The
microscopic transport studies in
Refs.~\cite{Cass99,Cass95C,Brat97,Ernst} for these systems have
given support for this interpretation. On the other hand,
 more conventional approaches that describe a melting of
the $\rho$-meson in the medium due to the strong hadronic coupling
(along the lines of Refs.~\cite{Rapp96,Peters}) have also been found
to be compatible with the early CERES
data~\cite{rapp5,Cass95C,CBRW97,Tserruya01}. This ambiguous
situation has been clarified to some extent in 2006 by the NA60
Collaboration since the invariant mass spectra for $\mu^+\mu^-$
pairs from In+In collisions at 158 A$\cdot$GeV  favored the `melting
$\rho$' scenario \cite{NA60}. Also, the later data from the CERES
Collaboration (with enhanced mass resolution)~\cite{CERES2} showed a
preference for the `melting $\rho$' picture.

The various models, which predict a change of the hadronic spectral
functions  in the (hot and dense) nuclear medium, may be classified
into two different categories: i) a broadening of the spectral
function or ii)  a mass shift of the vector mesons with density
and/or temperature. In view of many-body dynamics, both
modifications should be studied simultaneously as
well~\cite{MuellerSumRules,Tserruya:2009zt}. Thus we explore in the
present study three possible scenarios with respect to the low-mass
dilepton spectrum which essentially addresses all possible
properties of the $\rho$-meson: (1) a broadening of the $\rho$
spectral function, (2) a mass shift, and (3) a broadening plus a
mass shift.

We incorporate the effect of collisional broadening of the
vector-meson spectral functions (as in Refs.~\cite{BratKo99,GKC97}),
by using for the vector meson width
\begin{eqnarray}
\Gamma^*_V(M,|\vec p|,\rho_N)=\Gamma_V(M) + \Gamma_{coll}(M,|\vec
p|,\rho_N) . \label{gammas}
\end{eqnarray}
Here $\Gamma_V(M)$ is the total width of the vector mesons
($V=\rho,\omega$) in the vacuum.
The collisional width in (\ref{gammas}) is approximated as
\begin{eqnarray}
\Gamma_{coll}(M,|\vec p|,\rho_N) = \gamma \ \rho_N < v \
\sigma_{VN}^{tot} > \approx  \ \alpha_{coll} \ \frac{\rho_N}{\rho_0}
. \label{dgamma}
\end{eqnarray}
Here $v=|{\vec p}|/E; \ {\vec p}, \ E$ are the velocity, 3-momentum
and energy of the vector meson in the rest frame of the nucleon
current and $\gamma^2=1/(1-v^2)$. Furthermore, $\rho_N$ is the
nuclear density and $\sigma_{VN}^{tot}$ the meson-nucleon total
cross section.

In order to simplify the actual calculations for dilepton
production, the coefficient $\alpha_{coll}$ has been extracted in
the PHSD transport calculations from the vector-meson collision rate
in $In+In$ reactions at 158 A$\cdot$~GeV as a function of the
density $\rho_N$. In case of the $\rho$ meson the collision rate is
dominated by the absorption channels $\rho N \rightarrow \pi N$ or
$\rho N \rightarrow \Delta \pi \rightarrow \pi \pi N$. Also the
reactions $\rho +\pi \leftrightarrow a_1$ are incorporated. The
numerical results for $\Gamma_{coll}(\rho_N)$ then have been divided
by $\rho_N/\rho_0$ to fix the coefficient $\alpha_{coll}$ in
(\ref{dgamma}).  We obtain $\alpha_{coll} \approx 150$~MeV for the
$\rho$ and $\alpha_{coll} \approx 70$~MeV for $\omega$ mesons which
are consistent with Ref. \cite{Metag07}.  In this way the average
effects of collisional broadening are incorporated in accordance
with the transport calculations and allow for an explicit
representation of the vector-meson spectral functions versus the
nuclear density as demonstrated in Ref.~\cite{Brat08}.

In order to explore the observable consequences of vector meson mass
shifts at finite nuclear density, the in-medium vector meson pole
masses are modeled (optionally) according to the Hatsuda and Lee
\cite{H&L92} or Brown/Rho scaling \cite{BrownRho} as
\begin{eqnarray}
\label{Brown} M_0^*(\rho_N)= \frac{M_0} {\left(1 + \alpha {\rho_N /
\rho_0}\right)},
\end{eqnarray}
where $\rho_N$ is the nuclear density at the resonance decay
position $\vec r$; $\rho_0 = 0.16 \ {\rm fm}^{-3}$ is the normal
nuclear density and $\alpha \simeq 0.16$ for the $\rho$ and $\alpha
\simeq 0.12$ for the $\omega$ meson \cite{Metag07}. The
parametrization (\ref{Brown}) may be employed also at much higher
collision energies and one does not have to introduce a cut-off
density in order to avoid negative pole masses. Note that
(\ref{Brown}) is uniquely fixed by the 'customary' expression
$M_0^*(\rho_N) \approx M_0 (1 - \alpha \rho_N/\rho_0)$ in the low
density regime.

The spectral function of the vector meson $V$ for the mass $M$ at
baryon density $\rho_N$ is taken in the Breit-Wigner form:
\be A_V(M,\rho_N) \! = \! C_1{2\over \pi} {M^2
\Gamma_V^*(M\!,\rho_N) \over (M^2\!-\!M_{0}^{*^2}(\rho_N))^2 + (M
{\Gamma_V^*(M,\rho_N)})^2}\ . \label{spfunV} \ee
The factor $C_1$ is fixed by the normalization condition for
arbitrary $\rho_N$:
\begin{eqnarray}
\int_{M_{min}}^{M_{lim}} A_V(M,\rho_N) \ dM =1,
\label{SFnorma}\end{eqnarray} where $M_{lim}=2$~GeV is chosen as an
upper limit for the numerical integration. The lower limit for the
vacuum spectral function corresponds to the two-pion decay,
$M_{min}=2 m_\pi$, whereas for the in-medium collisional broadening
case $M_{min}=2 m_e \to 0$ with $m_e$ denoting the electron mass.
$M_0^*$ is the pole mass of the vector meson spectral function which
is $M_0^*(\rho_N=0)=M_0$ in vacuum, however, may be  shifted in the
medium for the dropping mass scenario according to Eq.
(\ref{Brown}).
The resulting spectral functions for the $\rho$ and $\omega$ meson
are displayed in Fig.~2 of Ref.~\cite{Brat08}.

With increasing nuclear density $\rho_N$ elastic and inelastic
interactions of the vector mesons shift strength to low invariant
masses. In the 'collisional broadening' scenario we find a dominant
enhancement of strength below the pole mass for the $\rho$ meson
while the $\omega$ meson spectral function is drastically enhanced
in the low- and high-mass region with density (on expense of the
pole-mass regime). In the 'dropping mass + collisional broadening'
scenario both vector mesons dominantly show a shift of strength to
low invariant masses with increasing $\rho_N$. Qualitatively similar
pictures are obtained for the $\phi$ meson but quantitatively
smaller effects are seen due to the lower effect of mass shifts and
a substantially reduced $\phi N$ cross section which is a
consequence of the $s\bar{s}$ substructure of the $\phi$ meson.

Note that, just as {the} HSD, {the} PHSD incorporates the {\em
off-shell propagation} for vector mesons -- according to
Ref.~\cite{Cass_off1}. In the off-shell transport{,} the hadron
spectral functions change dynamically during the propagation through
the medium and evolve towards the on-shell spectral function{s} in
the vacuum. The PHSD off-shell transport approach is particularly
suitable for investigating the different scenarios for the
modification of vector mesons in a hot and dense medium.  As
demonstrated in Ref.~\cite{Brat08}, the off-shell dynamics is
important for resonances with a rather long lifetime in {the} vacuum
but strongly decreasing lifetime in the nuclear medium (especially
$\omega$ and $\phi$ mesons) and also proves vital for the correct
description of dilepton decays  of $\rho$ mesons with masses close
to the two pion decay threshold. For a detailed description of the
off-shell dynamics we refer the reader to
Refs.~\cite{Cass_off1,Brat08,Bratkovskaya:2008bf,Linnyk:2011ee}.

\subsection{Multi-meson channels of dilepton production}

The dilepton excess yield in $In+In$ collisions at $160$~A$\cdot$GeV
incident energy for $M>1$~GeV/$c^2$ was found to be dominated by
partonic sources within the dynamical studies of Renk and
Ruppert~\cite{Renk:2008prc} as well as Dusling and
Zahed~\cite{Dusling}. On the other hand, the model of van Hees and
Rapp~\cite{RH:2008lp} suggests a dominance of hadronic sources
dubbed `4$\pi$ channels'. In order to clarify this question, we have
incorporated in {the} PHSD the `4$\pi$ channels' for dilepton
production on a microscopic level rather than assuming thermal
dilepton production and incorporating a parametrization for the
inverse reaction $\mu^+ + \mu^- \rightarrow 4 \pi's$ by employing
detailed balance as in {Refs.}~\cite{RH:2008lp,Santini:2011zw}.

By studying the electromagnetic emissivity (in the dilepton channel)
of the hot hadron gas, it was shown in
Ref{s}.~\cite{Song:1994zs,Gale:1993zj} that the dominating hadronic
reactions contributing to the dilepton yield at the invariant masses
above the $\phi$ peak are the two-body reactions, i.e.
$\pi+\rho$, $\pi+\omega$, $\rho+\rho$, $\pi+a_1$. This conclusion
was supported by the subsequent study in a hadronic relativistic
transport model~\cite{GLi}. Therefore, we implement the above listed
two-meson dilepton production channels in the PHSD approach. In
addition, some higher vector mesons ($\rho^\prime$ {\it etc.}) are
tacitly included by using phenomenological form factors adjusted to
data.

We determine the cross sections for the mesonic interactions with
dileptons in the final state using an effective Lagrangian approach,
following the works of Refs.~\cite{Song:1994zs,GLi}.
%
The dilepton production cross section is given by the product of a
form factor and the square of a scattering amplitude
\begin{equation} \label{cs}
\frac{d \sigma}{dt}=\frac{1}{64\pi s }\frac{1}{|p_{cm}|^2} |\bar
M|^2  |F(M)|^2,
\end{equation}
where
\begin{equation} \label{pcm}
p_{cm}=\sqrt{(s-(m_1+m_2)^2)(s-(m_1-m_2)^2)}/2\sqrt{s}
\end{equation}
is the center-of-mass momentum of the colliding hadrons with the
masses $m_1$, $m_2$, and $\bar{\vert{\cal M}\vert^2}$ can be written
as
\begin{equation} \label{LH}
\bar{\vert{\cal M}\vert^2}=4\left({4\pi\alpha\over q^2}\right)^2
L_{\mu\nu}H^{\mu\nu},
\end{equation}
with $q=p_1+p_2=p_3+p_4$ and the fine structure constant $\alpha$.
In (\ref{LH}), $L_{\mu\nu}$ is the leptonic tensor given by
\begin{equation}
L^{\mu\nu}=p_3^\mu p_4^\nu+p_4^\mu p_3^\nu-g^{\mu\nu} (p_3\cdot
p_4+m_l^2),
\end{equation}
while $H^{\mu\nu}$ is a hadronic tensor for the reaction.


The hadronic tensor $H^{\mu\nu}$ for the reaction $\pi^++\pi^-\to
e^++e^-$ is given by
\begin{equation}
H^{\mu\nu}=(p_2^\mu-p_1^\mu)(p_2^\nu-p_1^\nu),
\end{equation}
which leads to the well-known result for the $\pi \pi$ annihilation
cross section
\begin{equation} \label{cspipi}
\sigma_\pi(s)\!=\!{4\pi\alpha^2\over3s}\vert
F_\pi\vert^2\sqrt{1\!-\!{4 m_\pi^2\over s}} \left( 1\! -\!
\frac{4m_l^2}{M^2} \right) \left( 1\! +\! \frac{2m_l^2}{M^2}
\right)\!\!,\!\!
\end{equation}
where $M$ is the mass of the lepton pair, and $m_l$ is the mass of
the lepton. The electromagnetic form factor $|F_\pi (M)|^2$ plays an
important role in this process, providing empirical support for the
vector meson dominance: the pion electromagnetic form factor is
dominated by the $\rho (770)$ meson. In Ref. \cite{gale87}, Gale and
Kapusta proposed the form,
\begin{eqnarray}\label{gk87}
|F_\pi (M)| ^2 = {m_r^4\over (M^2-m_r^{\prime 2} )^2 +(m_r\Gamma
_r)^2},
\end{eqnarray}
where $m_r=0.775$~GeV, $m_r^\prime = 0.761$~GeV, and $\Gamma
_r=0.118$~GeV.

\begin{figure}
\centering
\subfigure[]{ \label{piR}
\includegraphics[width=0.46\textwidth]{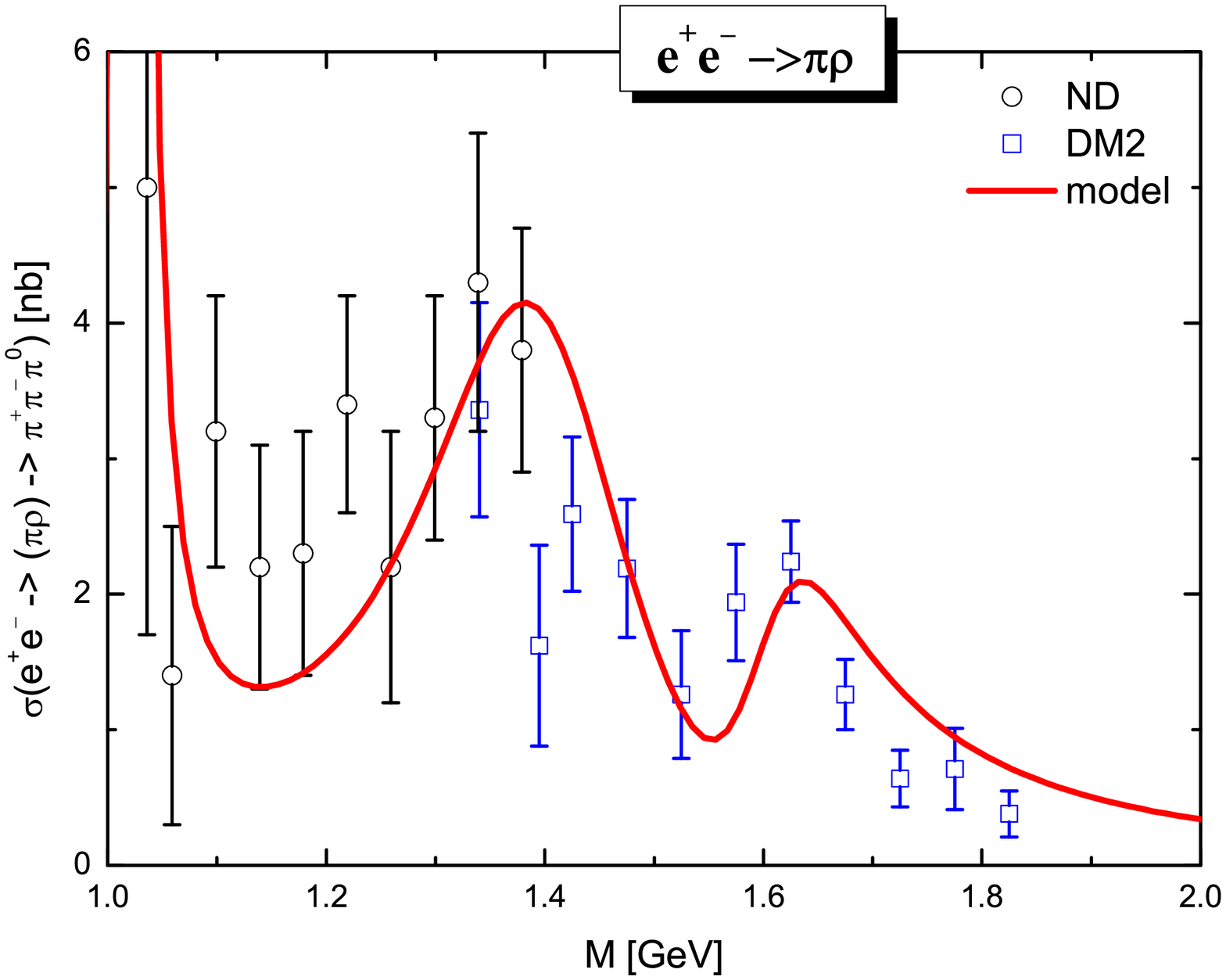} }
\subfigure[]{ \label{piO}
\includegraphics[width=0.47\textwidth]{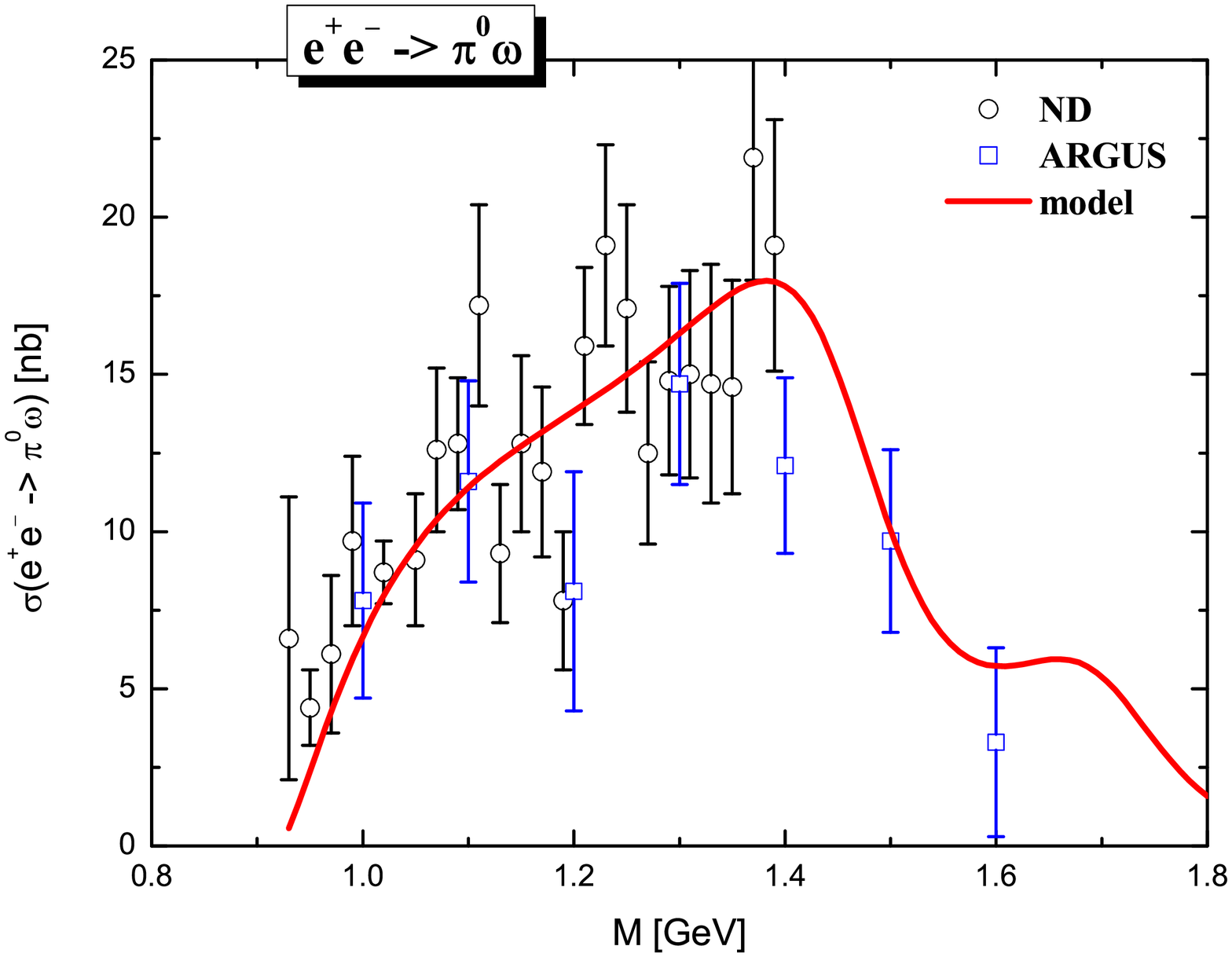} }
\subfigure[]{ \label{pia1_RR}
\includegraphics[width=0.47\textwidth]{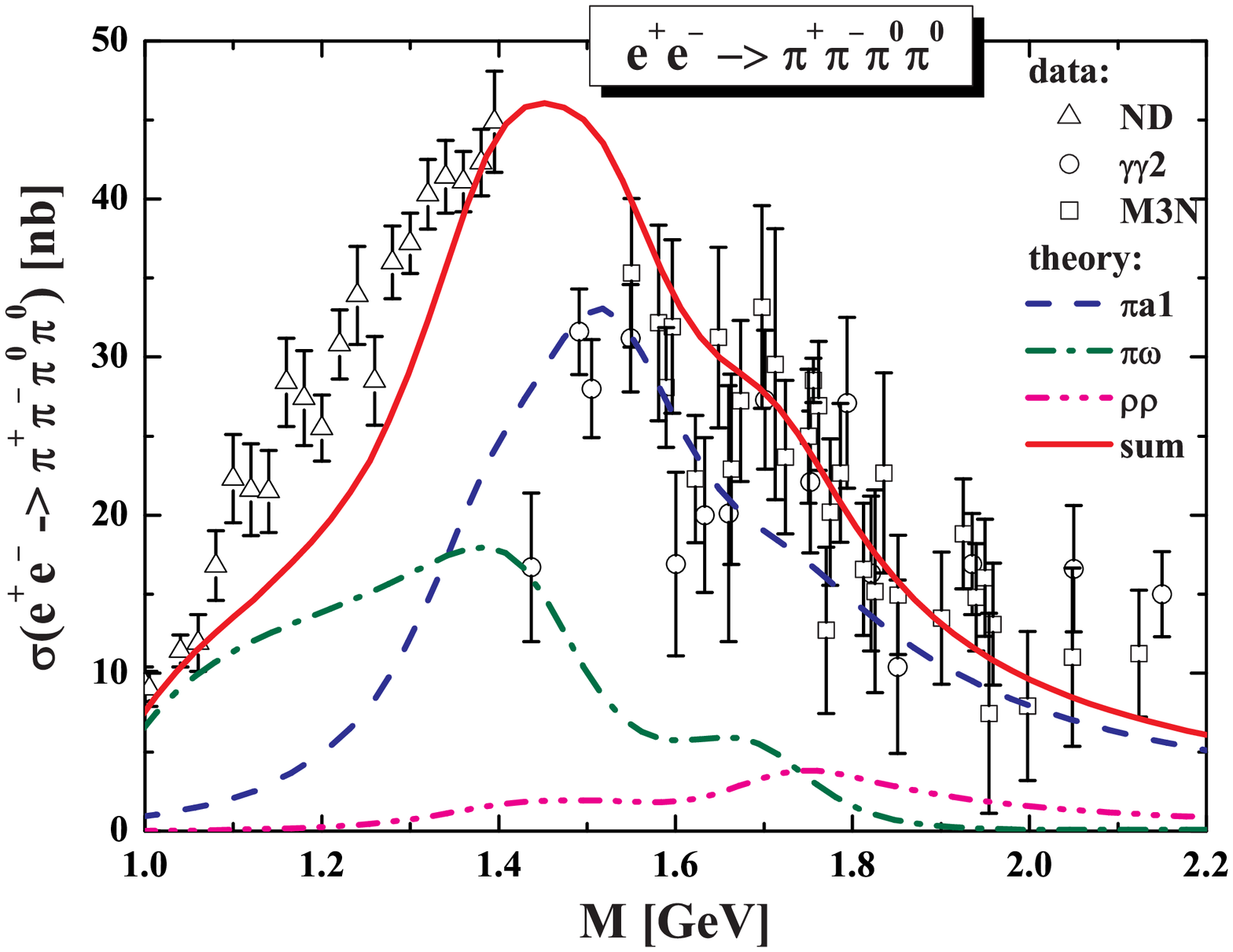} }
\caption{(color online) a) Cross sections for the reactions
$e^+e^-\to \pi+\rho$ (a) and $e^+e^-\to\pi+\omega$ (b) in our model
versus the experimental data. Panel (c) gives the measured cross
section of the $e^+e^-\to \pi^+\pi^-\pi^0\pi^0$ reaction versus the
sum of the model cross sections for $e^+e^-\to \pi+\omega$,
$e^+e^-\to \pi+a_1$ and $e^+e^-\to \rho+\rho$.} \label{ee}
\end{figure}


According to Ref.~\cite{haglin95}, the cross section for $\pi\rho$
annihilation is given by
\be \label{cspiro} \sigma (\pi^+\rho^- \rightarrow l{\bar l}) =
{2\pi \alpha ^2 p_{cm} \over 9 M} |F_{\pi\rho} |^2
\Big(1\!-\!{4m_l^2\over M^2}\Big) \Big(1\!+\!{2m_l^2\over M^2}\Big).
\ee
Note that the cross section (\ref{cspiro}) is evaluated in the
narrow-width approximation  for illustration purposes only. This
simplification is not used in the actual transport calculation.
The electromagnetic form factor $|F_{\pi\rho} (M)|^2$ can then be
determined by analyzing the experimental data for $e^+e^-\rightarrow
\pi^+\pi^-\pi^0$. In Ref.~\cite{haglin95}, three isoscalar vector
mesons, $\phi (1020)$, $\omega (1420)$, and $\omega (1670)$ were
found to be important in order to fit the experimental data, namely,
\begin{eqnarray}
F_{\pi\rho} (M) = \sum _V \left({g_{V\pi\rho}\over g_V}\right)
{e^{i\phi _V} m_V^2\over (m_V^2-M^2) - i m_V \Gamma _V}.
\end{eqnarray}
Here the summation runs over the three vector mesons listed above.
While the coupling constants $g_\phi$ and $g_{\phi\pi\rho}$ can be
determined from the measured widths, the coupling constants for
other two mesons and the relative phases were determined by a fit to
the experimental data of Refs.~\cite{aul86,bal87}. These coupling
constants were extracted from the latest data of the DM2
collaboration~\cite{dm2piro} and the ND collaboration~\cite{nd}. The
parameters are listed in Ref.~\cite{haglin95}. The comparison of the
fit to the experimental data is shown in Fig.~\ref{piR}.


The cross section for lepton pair production in pion-omega
annihilation is given by~\cite{GLi}
\be \sigma (\pi^0\omega \rightarrow l{\bar l}) = {4\pi \alpha ^2
p_{cm} \over 9 M} |F_{\pi\omega} |^2 \Big(1-{4m_l^2\over M^2}\Big)
\Big(1+{2m_l^2\over M^2}\Big). \ \ \ee
The form factor can be parametrized in terms of three isovector
$\rho$-like vector mesons, $\rho (770)$, $\rho (1450)$, and $\rho
(1700)$,
\begin{eqnarray}
F_{\pi\omega} (M) = \sum _V \left({g_{V\pi\omega}\over g_V}\right)
{e^{i\phi _V} m_V^2\over (m_V^2-M^2) - i m_V \Gamma _V}.
\end{eqnarray}
Here the summation runs over the three $\rho$-like resonances listed
above. The parameters used are
$m_{r1}=0.77$~GeV, $m_{r2}=1.45$~GeV, $m_{r3}=1.7$~GeV,
$\Gamma_{r1}=0.118$~GeV, $\Gamma_{r2}=0.25$~GeV,
$\Gamma_{r3}=0.22$~GeV, $A_{r1}=0.85$, $A_{r2}=-0.077$,
$A_{r3}=0.034$, where $A_{V}=(g_{V\pi\omega}/g_V)\exp\{i\phi_V\}$.
The comparison with the experimental data of the ND~\cite{nd} and
ARGUS collaborations~\cite{argus} is shown in Fig.~\ref{piO}.


Additionally, we consider the reactions $\pi a_1 \rightarrow l{\bar
l}$ and $\rho \rho \rightarrow l{\bar l}$, which are effectively
four-pion processes. Using the Lagrangian for the $\pi a_1$
interaction
\begin{equation}
{\cal L}_{\pi a_1 \gamma*} = g e a^\mu [(\partial_\nu A
_\mu)(\partial^\nu \pi) - (\partial_\mu A^\nu)(\partial_\nu \pi)],
\end{equation}
one obtains for the cross section of the $\pi a_1 \rightarrow l{\bar
l}$ process,
\begin{eqnarray}
\sigma (\pi a_1 \rightarrow l{\bar l}) & = & \frac{\pi \alpha^2 g^2
M}{3 p_{cm}} \left(1-{4m_l^2\over M^2}\right) \left(1+{2m_l^2\over
M^2}\right) \nn
&& \hspace{-1.8cm} \times \left\{ \frac{1}{4}
\left(1-\frac{m_{a1}^2}{M^2}\right)
   \left[  1 + \frac{2}{m_{a1}^2}\left(\frac{5p_{cm}^2}{12}+\frac{m_{a1}^2}{2}\right)
   \right] \right. \nn
&& \hspace{-1.3cm} + \left(1-\frac{m_{a1}^2}{M^2}\right) \left[
 -  \frac{1}{2}\left(1-\frac{m_{a1}^2}{M^2}\right)   +  \frac{\sqrt{p_{cm}^2+m_{\pi}^2}}{M}
 \right.
 \nn
&& \hspace{-0.65cm}
-\frac{M^2}{2m_{a1}^2}\left(1+\frac{m_{a1}^2}{M^2}\right) \left(
 \frac{p_{cm}^2}{6M^2}-\frac{1}{2}\left(1-\frac{m_{a1}^2}{M^2}\right) \right.\nn
&& \hspace{1.1cm} \left. \left.
 +\frac{\sqrt{p_{cm}^2+m_{a1}^2}\sqrt{p_{cm}^2+m_{\pi}^2}}{M^2}
\right) \right] \nn
&& \hspace{-1.3cm} \left.
+\frac{5p_{cm}^2}{6M^2}\left[\frac{(M^2+m_{a1}^2)^2}{4m_{a1}^2M^2}-1\right]
\right\} |F_{\pi a_1}|^2,
\end{eqnarray}
where the value of the coupling constant $g=(g_\rho/f_\rho)$ is
adjusted so that the experimentally measured radiative decay widths
are reproduced.


We obtain the hadronic tensor $H^{\mu\nu}$ for the reaction
$\rho^+\rho^-\to e^+e^-$ by generalizing the formula of
Ref.~\cite{Song:1994zs} to explicitly take into account the broad
spectral functions of the colliding $\rho$-mesons:
\begin{eqnarray} \label{Hroro}
H^{\mu\nu}&=&h_\rho^{\mu\alpha\beta} h_{\rho\,\alpha\beta}^\nu
-h_\rho^{\mu\alpha\beta}p_{1\,\beta}
 h_{\rho\,\alpha}^{\nu\gamma}p_{1\,\gamma}/m_{\rho 1}^2 \nn
&&-h_\rho^{\mu\alpha\beta}p_{2\,\alpha}
 h_{\rho\,\beta}^{\nu\gamma}p_{2\,\gamma}/m_{\rho 2}^2 \nn
&& +h_\rho^{\mu\alpha\beta}p_{1\,\beta}p_{2\,\alpha}
 h_\rho^{\nu\gamma\delta}p_{2\,\gamma}p_{1\,\delta}/(m_{\rho 1}^2 m_{\rho 2}^2),
\end{eqnarray}
with
\begin{equation}
h_\rho^{\mu\alpha\beta}=(p_2^\mu-p_1^\mu)g^{\alpha\beta}
                  +(q^\alpha-p_2^\alpha)g^{\beta\mu}
                  +(p_1^\beta-q^\beta)g^{\mu\alpha}.
\end{equation}
In this case, the hadronic tensor depends on (generally different)
masses of the colliding particles $m_{\rho 1}$ and $m_{\rho 2}$. In
the actual transport calculations, $m_{\rho i}$ are distributed
according to the dynamical spectral functions. Using
(\ref{cs})-(\ref{LH}) and (\ref{Hroro}) we obtain the following
cross section as a function of $M\!$, $m_{\rho 1}$ and $m_{\rho 2}$
\bea
\sigma (\rho\rho\to l^+l^-)&=&
\frac{\pi\alpha^2|F_{\rho\rho}|^2}{120m_{\rho 1}^2m_{\rho 2}^2M^5
p_{cm}} \nn
&& \hspace{-2.5cm}
\times \left\{ 9m_{\rho 1}^8
+ 18 m_{\rho 1}^6(3m_{\rho 2}^2-2M^2) \right. \nn
&& \hspace{-2.1cm}
+ (m_{\rho 2}^2-M^2)^2\left[819m_{\rho 2}^4+632m_{\rho
2}^2M^2-11M^4\right] \nn
&& \hspace{-2.1cm}
- 2m_{\rho 1}^2\left[363m_{\rho 2}^6+32m_{\rho 2}^4M^2+327m_{\rho
2}^2M^4-2M^6\right] \nn
&& \hspace{-2.1cm} \left.
+ m_{\rho 1}^4\left[-156m_{\rho 2}^4+266m_{\rho 2}^2M^2+34M^4\right]
 \right\},
\eea
which reduces in the narrow width approximation to
\bea
\hspace{-0.7cm} \sigma (\rho\rho\to l^+l^-)&=&
\frac{\pi\alpha^2|F_{\rho\rho}|^2}{60m_\rho
^4M^3\sqrt{M^2-4m_\rho^2}} \nn
&& \hspace{-2.2cm}
\times \left( 840m_\rho^6+1076m_\rho^4M^2-658m_\rho^2M^4+11M^6
\right)\!. \
\eea

The form factors $|F_{\pi a_1}|^2$ and $|F_{\rho\rho}|^2$ can be
determined by analyzing the $e^+e^-\rightarrow \pi^+\pi^-\pi^+\pi^-$
and $e^+e^-\rightarrow \pi^+\pi^- \pi^0 \pi^0$ data. We determine
$|F_{\pi a_1} (M)|^2$ from experimental data for $e^+e^-\rightarrow
\pi^+\pi^-\pi^+\pi^-$ from the $\gamma \gamma 2$ collaboration
\cite{rr2}, the M3N collaboration \cite{m3nc}, and the ND
collaboration \cite{nd}. Further constraints on $|F_{\pi a_1}|^2$
and the determination of $|F_{\rho\rho}|^2$ were provided by the
experimental data for $e^+e^-\rightarrow \pi^+\pi^-\pi^0\pi^0$,
which can come from $\pi\omega $, $\pi a_1 $ and $\rho\rho$
intermediate states. Our form factors are
\be
F_{\pi a_1} (M) = \sum _V \left({g_{V\pi a_1}\over g_V}\right)
{e^{i\phi _V} m_V^2\over (m_V^2-M^2) - i m_V \Gamma _V}
\ee
with
$m_{r1}=0.77$~GeV, $m_{r2}=1.45$~GeV, $m_{r3}=1.7$~GeV,
$\Gamma_{r1}=0.118$~GeV, $\Gamma_{r2}=0.25$~GeV,
$\Gamma_{r3}=0.235$~GeV, $A_{r1}=0.05$, $A_{r2}=0.58$,
$A_{r3}=0.027$,
and
\be
F_{\rho\rho} (M) = \sum _V \left({g_{V\rho\rho}\over g_V}\right)
{e^{i\phi _V} m_V^2\over (m_V^2-M^2) - i m_V \Gamma _V}
\ee
with
$m_{r1}=0.77$~GeV, $m_{r2}=1.45$~GeV, $m_{r3}=1.7$~GeV,
$\Gamma_{r1}=0.118$~GeV, $\Gamma_{r2}=0.237$~GeV,
$\Gamma_{r3}=0.235$~GeV, $A_{r1}=0.05$, $A_{r2}=0.05$,
$A_{r3}=0.02$.
The comparison to the data is shown in Fig.~\ref{pia1_RR}.

Let us summarize that in order to fix the form factors in the cross
sections for dilepton production by the interaction of $\pi+\rho$,
$\pi +\omega$, $\rho+\rho$ and $\pi a_1$, we use the measurements in
the detailed-balance related channels: $e^+e^-\to \pi+\rho$,
$e^+e^-\to \pi +\omega$, $e^+e^-\to \rho+\rho$, and $e^+e^-\to
\pi+a_1$. Note that we fitted the form factors while taking into
account the widths of the $\rho$ and $a_1$ mesons in the final state
by convoluting the cross sections with the (vacuum) spectral
functions of these mesons in line with Ref.~\cite{Song:1995wy}
(using the parametrizations of the spectral functions as implemented
in HSD and described in~\cite{Bratkovskaya:2008iq}). In
Fig.~\ref{ee} we present the resulting cross sections, which are
related by detailed balance to the ones we implemented into PHSD.


\begin{figure*}
\centering \subfigure[]{ \label{Thermal1}
\resizebox{0.48\textwidth}{!}{%
 \includegraphics{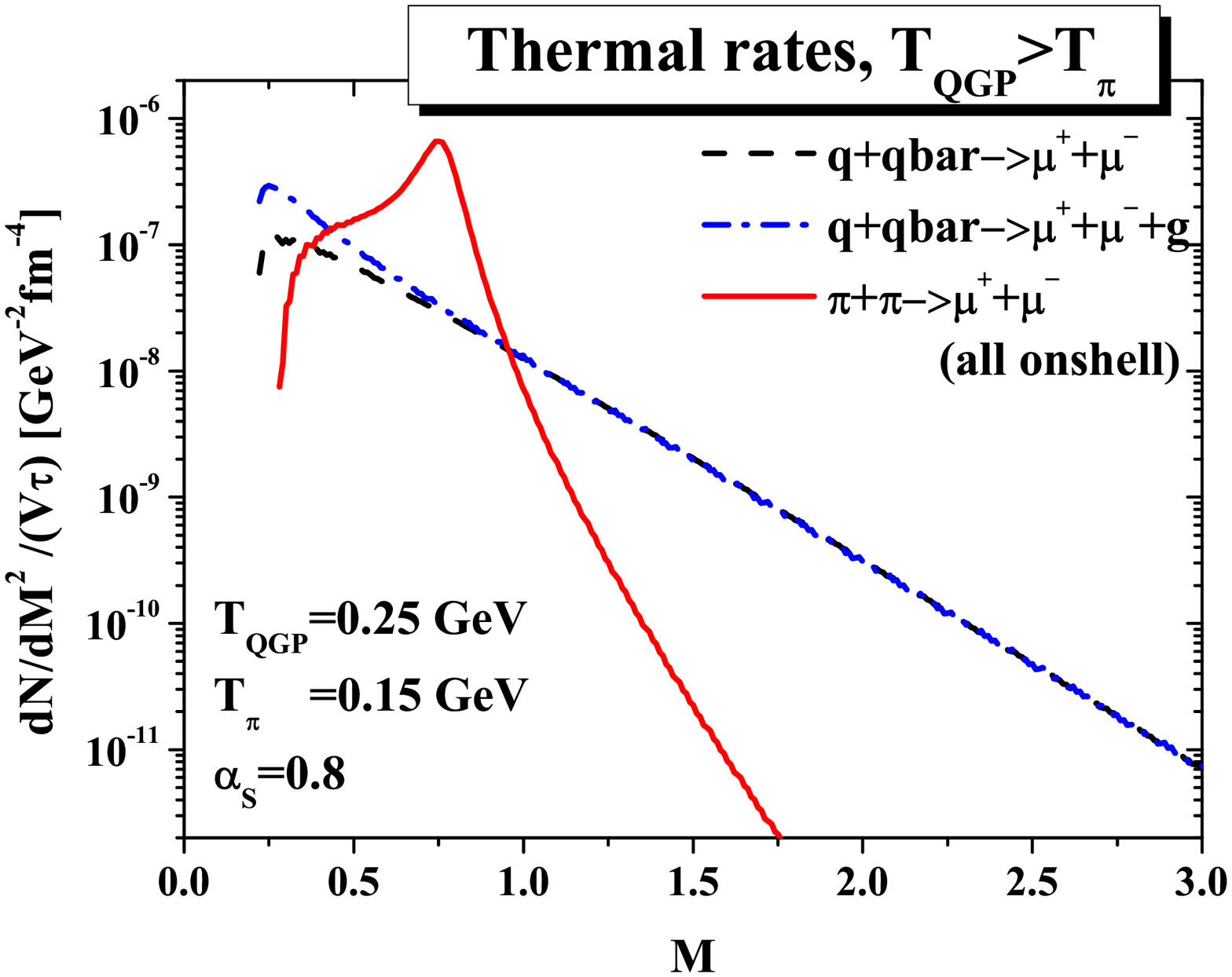}
} } \subfigure[]{ \label{Thermal2}
\resizebox{0.48\textwidth}{!}{%
 \includegraphics{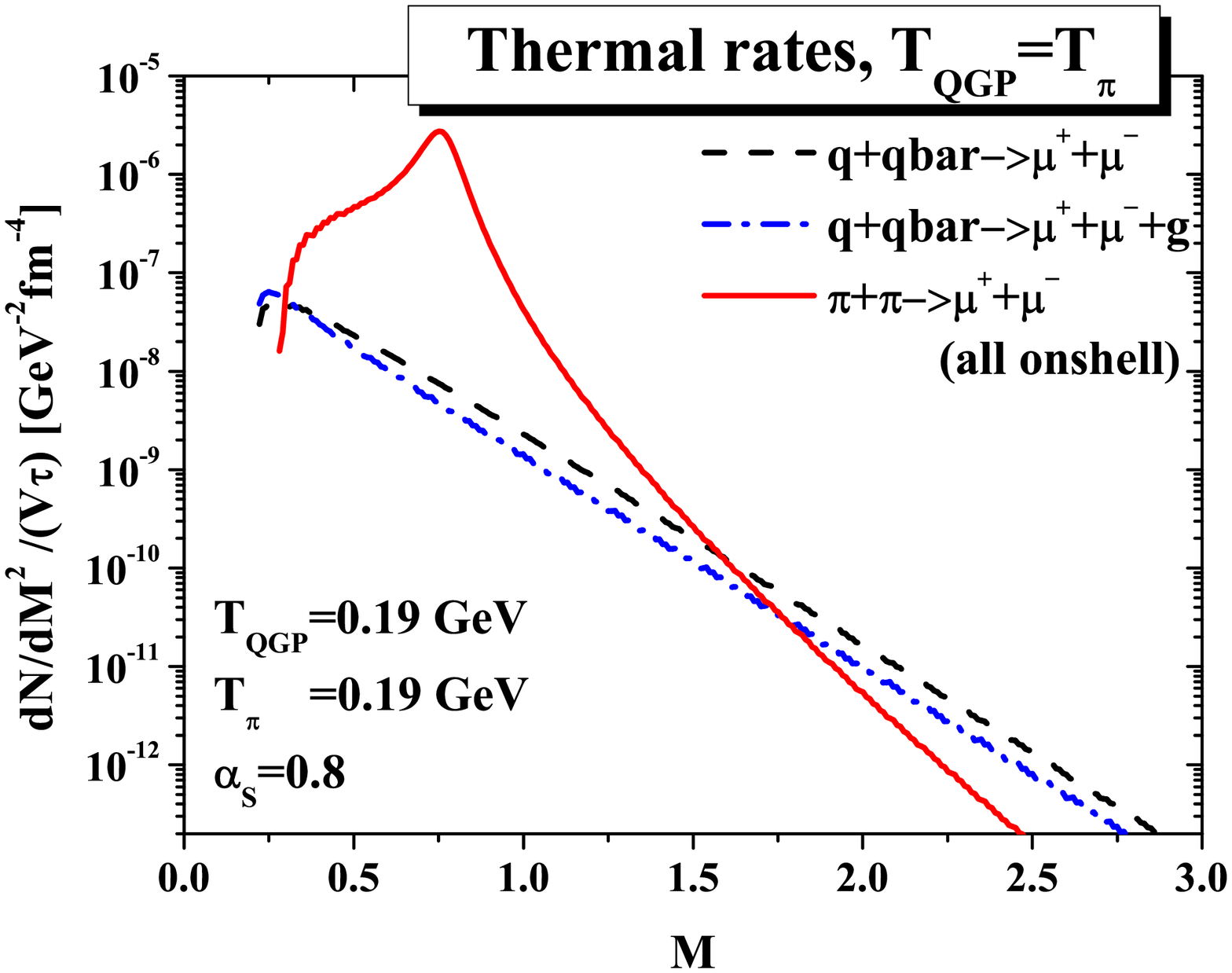}
} } \caption{(color online) Rates of dileptons created in $q+\bar q$
and $\pi+\pi$ annihilations within a thermalized gas of quarks with
temperature $T_{QGP}$ and pions with temperature $T_\pi$.  (a)
$T_{QGP}=250$~MeV, $T_\pi=150$~MeV; (b) $T_{QGP}=T_\pi=190$~MeV.
\label{Thermal} }
\end{figure*}

\section{Dilepton rates in thermal equilibrium}
\label{section.thermal}

Before proceeding to the results of the transport calculations and
the comparison to data, we dedicate this section to a study of the
dilepton spectrum qualitatively in a thermal model.
In Fig.~\ref{Thermal}, the dilepton production rates in thermal
equilibrium are presented. We assume here that the system evolves
through a thermalized system of quark in the hot initial stage of
the heavy-ion collision and through the state of a high-density
hadron gas in the later phase of the collision.

The  main elementary process of dilepton production in a hadron gas
is the pion annihilation into dileptons, mediated through vector
meson dominance by the rho meson ($\pi+\pi\to\rho\to\gamma^*\to l^+
+l^-$) and controlled by the pole at the rho mass of the pion
electromagnetic form factor. For the pion annihilation, we use the
standard cross section as, e.g., in Ref.~\cite{Song:1994zs} and the
Breit-Wigner form factor with the pole mass and width of the $\rho$
meson.

In the partonic sector, the main sources of the dileptons are the
reactions of quark-antiquark annihilation with the production of the
virtual photon. Considering the temperatures and baryon densities
relevant for the SPS energies, we expect in PHSD the contribution of
the processes involving gluons to be small compared to the leading
$q+\bar q$ mechanism of dilepton production (note, however, that at
higher energies, such as those of RHIC and LHC, gluons can play an
important role in the dilepton production~\cite{Lin:1999uz}). For
the calculation of the QGP yield in the qualitative analysis of this
section, the most simple perturbative QCD cross sections are used
for the processes $q+\bar q\to l^+l^-$ and $q+\bar q\to g+l^+l^-$,
assuming $\alpha_S=0.8$.

Thus we plot the dilepton yields from the reactions $\pi+\pi$ and
$q+\bar q$, where the pions and quarks have in general different
temperatures $T_\pi$ and $T_{QGP}$. The space-time volumes of the
two phases are assumed to be approximately equal. In
Fig.~\ref{Thermal1}, the gas of pions is assumed to have the
temperature $T_{\pi}=150$~MeV, while the gas of quarks the
temperature $T_{QGP}=250$~MeV. In Fig.~\ref{Thermal2}, {we have}
$T_{QGP}=T_\pi=190$~MeV.

It has been originally suggested that a `window' for observing
dileptons from the plasma exists in the invariant mass region
between the $\phi$ and $J/\Psi$ peaks~\cite{Shuryak:1978ij}. This is
supported by the results shown in both Figs.~\ref{Thermal1} and
\ref{Thermal2}. However, we see in Fig.~\ref{Thermal1} another
region, i.e. $M<0.5$~GeV, in which the $q+\bar q$ annihilation is
compatible or even larger than the radiation from the $\pi+\pi$
annihilation; the contribution of the two-to-two process $q+\bar
q\to g+l^+l^-$ is especially important. The dominance of the thermal
yield from quark interactions at masses below $\approx 0.5$~GeV is
in agreement with the conclusions
of~\cite{Gallmeister:1999dj,Gallmeister:2000ra}. The transport model
results of the next section will clarify which of the equilibrium
scenarios presented in Fig.~\ref{Thermal} -- (a) or (b) -- {gives a}
closer resembl{ance to} the channel decomposition of the dilepton
production within a microscopic simulation.

Of course, the observation of the QGP channels at low mass is
possible only after the dilepton yield from the $\pi$-, $\eta$- and
$\omega$-Dalitz decays is removed. Another word of caution is in
place here, because in Figs.~\ref{Thermal} the vacuum properties of
the $\rho$-meson have been used in plotting the $\pi+\pi$
contribution, whereas the $\rho$-meson properties are expected to be
modified in medium. The modification of the $\rho$ will change the
size of the new, low mass window of the QGP observation (cf. next
section).


\begin{figure*}
\includegraphics[width=\textwidth]{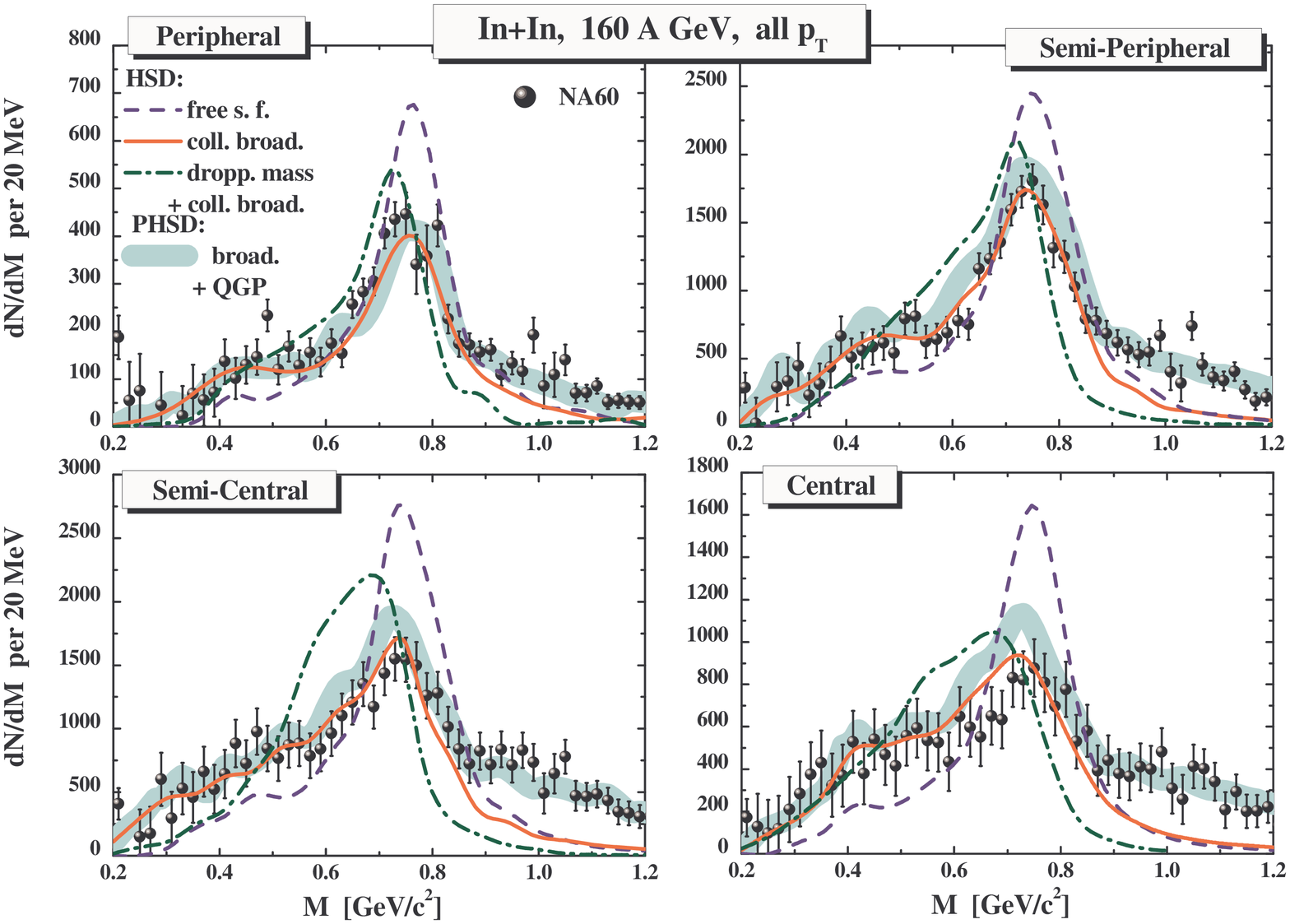}
\caption{(color online) The HSD results for the mass differential
dilepton spectra from $In + In$ collisions at 158 A$\cdot$GeV in
comparison to the excess mass spectrum from NA60~\cite{NA60}. The
actual NA60 acceptance filter and mass resolution have been
incorporated~\cite{Sanja}.  The solid lines show the HSD results for
a scenario including the collisional broadening of the $\rho$-meson
whereas the dashed lines correspond to calculations with  'free'
$\rho$ spectral functions for reference. The dash-dotted lines
represent the HSD calculations for the 'dropping mass + collisional
broadening' model. The (blue) bands represent the  PHSD results
incorporating direct dilepton radiation from the QGP in addition to
a broadened $\rho$-meson.} \label{wtf}
\end{figure*}

\section{Results and comparison to data}
\label{section.results}

Let us first note that the bulk properties of heavy-ion reactions at
the top SPS energy, such as the number of charged particles, as well
as their rapidity, $p_T$ and transverse energy distributions, were
rather well described by PHSD; we refer to Ref.~\cite{CasBrat} for
an extended and detailed comparison to the data. As the lQCD
equation of state employed {here has} a crossover transition, the
PHSD calculations show a rather long QGP phase in central $In+In$
collisions at 158 A$\cdot$GeV (cf. Fig.~10 of Ref.~\cite{CasBrat})
with the partonic degrees of freedom dominating for about 3~fm/c.
Also, the elementary $pp$ channel is well under control in PHSD as
has been demonstrated in Ref.~\cite{Bratkovskaya:2008bf}.

Previously, by employing the HSD approach to the low mass dilepton
production in relativistic heavy-ion collisions, it was shown in
Ref.~\cite{Bratkovskaya:2008bf} that the NA60 Collaboration data for
the invariant mass spectra of $\mu^+\mu^-$ pairs from In+In
collisions at 158 A$\cdot$GeV favored the 'melting $\rho$'
scenario~\cite{NA60}. Also the data from the CERES
Collaboration~\cite{CERES2} showed a preference for the 'melting
$\rho$' picture. For other vector mesons ($\omega, \phi$), the
effects are relatively small, since, due to their much longer
lifetimes, $\omega$ and $\phi$ decay predominantly outside the
medium after regaining the vacuum properties.

As we see in Fig.~\ref{wtf}, the current calculation in the PHSD
approach confirms the earlier finding in the hadronic model that the
NA60 data favor the scenario of the in-medium broadening of vector
mesons. A comparison of the transport calculations to the data of
the NA60 Collaborations points towards a 'melting' of the
$\rho$'-meson at high densities, i.e. a broadening of the vector
meson's spectral function in line with the findings by Rapp and
collaborators~\cite{Rapp}. No pronounced mass shift of the vector
mesons is visible in the data. Thus the experimental results suggest
that  the approach to the chiral transition proceeds through
broadening, and eventually melting, of the resonances rather than by
dropping masses. On the other hand, a closer inspection of Fig.~\ref{wtf} shows that
the conventional hadronic sources do not match the measured yield at
invariant masses above about 1~GeV/$c^2$, while the yield at masses
close to 1~GeV is reproduced by taking into account the dilepton
production channels in the QGP.

\begin{figure}
\includegraphics[width=0.5\textwidth]{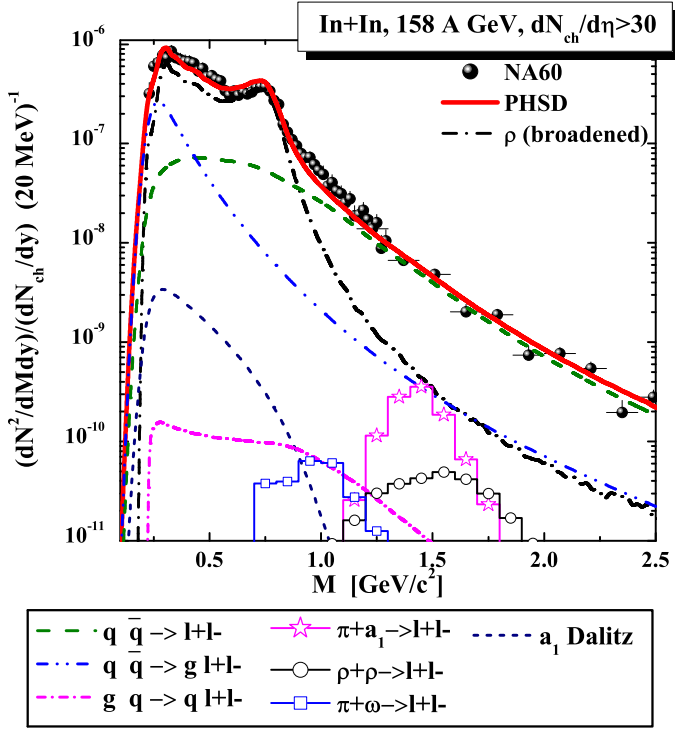}
\caption{(color online) Acceptance corrected mass spectra of excess
dimuons from $In+In$ at 158~AGeV integrated over $p_T$ in
$0.2<p_T<2.4$~GeV from PHSD compared to the data of
NA60~\cite{Arnaldi:2008er}. The dash-dotted line shows the dilepton
yield from the in-medium $\rho$ with a broadened spectral function,
the dashed line presents the yield from the $q+\bar q$ annihilation,
the dash-dot-dot line gives the contribution of the gluon
Bremsstrahlung process ($q\bar q\to g l^+l^-$), while the solid line
is the sum of all contributions. For the description of the other
lines, which correspond to the non-dominant channels, we refer to
the figure legend.} \label{NA60_AC}
\end{figure}

The NA60 collaboration has recently published acceptance corrected
data with subtracted charm contribution~\cite{Arnaldi:2008er}. In
Fig.~\ref{NA60_AC} we present PHSD results for the dilepton spectrum
excess over the known hadronic sources as produced in $In+In$
reactions at 158~A$\cdot$GeV compared to the acceptance corrected
data. We find here that the spectrum at invariant masses in the
vicinity of the $\rho$ peak is well reproduced by the $\rho$ meson
yield, if a broadening of the meson spectral function in the medium
is assumed, while the partonic sources account for the yield at high
masses.

One concludes from Fig.~\ref{NA60_AC} that the measured spectrum for
$M>1$~GeV is dominated by the {\it partonic} sources. Indeed, the
domination of the radiation from {the} QGP over the hadronic sources
in PHSD is related to a rather long -- of the order or 3 fm/c --
evolution in the partonic phase (in co-existence with the space-time
separated hadronic phase) on one hand, cf. Fig.~10 of
Ref.~\cite{CasBrat}, and the rather high initial energy densities
created in the collision on the other hand, cf. Fig.~6 of
Ref.~\cite{Linnyk:2008hp}.

In addition, we find from Fig.~\ref{NA60_AC} that in PHSD the
partonic sources have a considerable (about 30\%) contribution to
the dilepton yield at $M<0.6$~GeV. The yield from the two-to-two
process $q+\bar q\to g+l^+l^-$ is especially important. This
conclusion from the microscopic calculation is in qualitative
agreement with the conclusion of an early (more schematic)
investigation in Ref.\cite{Alam:2009da}.

Recalling the illustrative study of dilepton rates in thermal
equilibrium in section~\ref{section.thermal}, we observe that the
non-equilibrium microscopic simulation within the PHSD transport
approach qualitatively implies a situation in which the initial
partonic phase has temperatures of the order of
$T_{QGP}\approx250$~MeV and the hadron gas in the subsequent
evolution a temperature {$T_\pi\approx$150}~MeV (assuming
thermalization and that their evolution is approximately as long). A
model scenario, in which the temperatures of the partonic and
hadronic phases are equal for an extended period of space-time
($T_\pi=T_{QGP}=190$~MeV) is not supported by the microscopic
simulations.

\begin{figure}[t]
\includegraphics[width=0.5\textwidth]{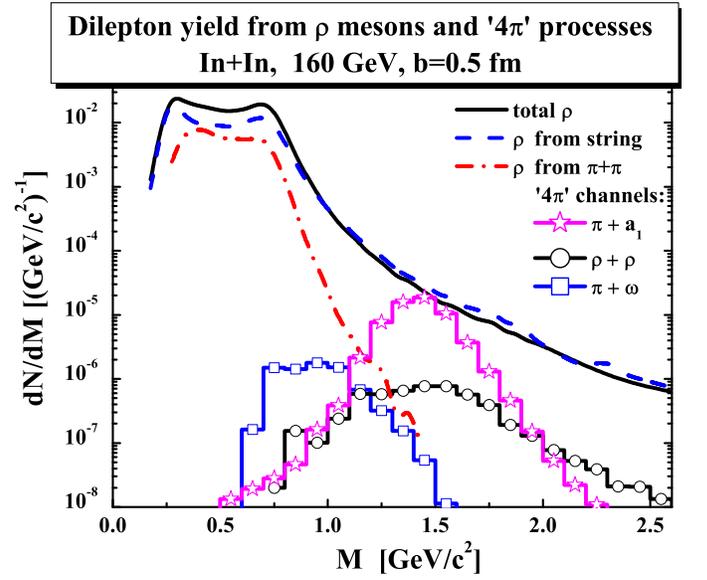}
\caption{(color online) Dilepton radiation from $\rho$-mesons of
different origins in PHSD from central $In+In$ collisions at
158~A$\cdot$GeV compared to the contributions from the `4$\pi$'
processes ($a_1+\pi$, $\pi+\omega$, $\rho+\rho $). The direct
$\rho$-mesons produced in mesonic and baryonic strings are given by
the dashed line and the `thermal' $\rho$-mesons produced in
$\pi+\pi$ annihilations by the dash-dotted line. The contributions
of the `$4\pi$' processes are shown by the lines with symbols: the
$\pi+a_1\to l^+l^-$ process is displayed by the line with stars,
$\pi+w\to l^+l^-$ by the line with squares and $\rho+\rho\to l^+l^-$
by the line with circles.} \label{Rhos}
\end{figure}

In order to elucidate the relative importance of the different {\it
hadronic} sources of the excess dileptons in the heavy-ion
collisions at top SPS energies, we show in Fig.~\ref{Rhos} the
channel decomposition of the main hadronic contributions to the
dilepton rates in central $In+In$ collisions at $158$~A$\cdot$GeV
integrated over rapidity and $p_T$. In particular, the dilepton
yield from the decays of the $\rho$-mesons (solid line) is
dominantly composed of two channels: the direct $\rho$-mesons
produced in mesonic and baryonic strings (dashed line) and the
`thermal' $\rho$-mesons produced in  $\pi+\pi$ annihilations
(dash-dotted line). For comparison, the contributions of the
`$4\pi$' processes are shown by the lines with symbols: the
$\pi+a_1\to l^+l^-$ process is displayed by the line with stars,
$\pi+w\to l^+l^-$ by the line with squares and $\rho+\rho\to l^+l^-$
by the line with circles. We find that the dilepton yield from the
decays of the `thermal' $\rho$-mesons falls exponentially at high
masses. The contributions from the `$4\pi$' processes start
dominating over the yield from the `thermal' $\rho$ decays at
$M\approx1$~GeV. We further confirm in PHSD that at $M>1$~GeV the
contribution of the $\pi+a_1$ process is the highest among the
secondary mesonic sources of the dileptons, as was first noted by
the authors of Ref.~\cite{Song:1994zs}. On the other hand, in
contrast to the `thermal' $\rho${'}s, the direct $\rho$-mesons
produced in the string decays (following the initial hard
collisions) exhibit a power-law tail at masses above 1~GeV and,
consequently, dominate the overall dilepton spectrum of hadronic
origin for $M>1$~GeV.

\begin{figure}
\includegraphics[width=0.5\textwidth]{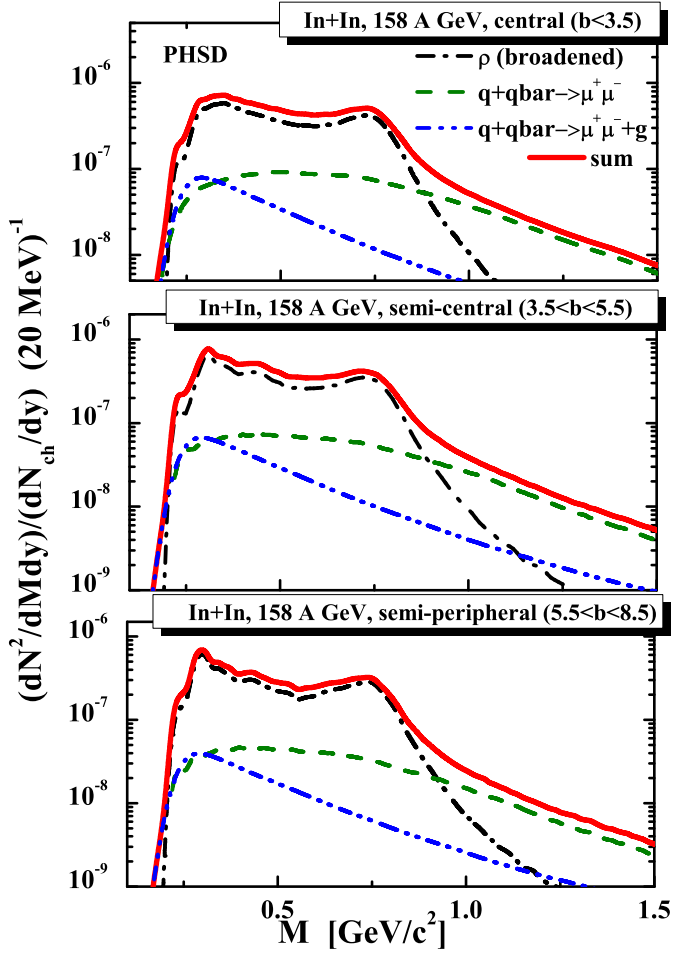}
\caption{(color online) Mass spectra of excess dimuons from $In+In$
at 158~AGeV for $0.2<p_T<2.4$~GeV and $3<\eta<4.2$ from PHSD for
different centrality bins. The dash-dotted, dashed, and solid lines
show, respectively, the dilepton yield from the in-medium $\rho$
with a broadened spectral function, the dilepton yield from the
$q+\bar q$ annihilation and the sum of them.} \label{centrality}
\end{figure}

\begin{figure}
\includegraphics[width=0.5\textwidth]{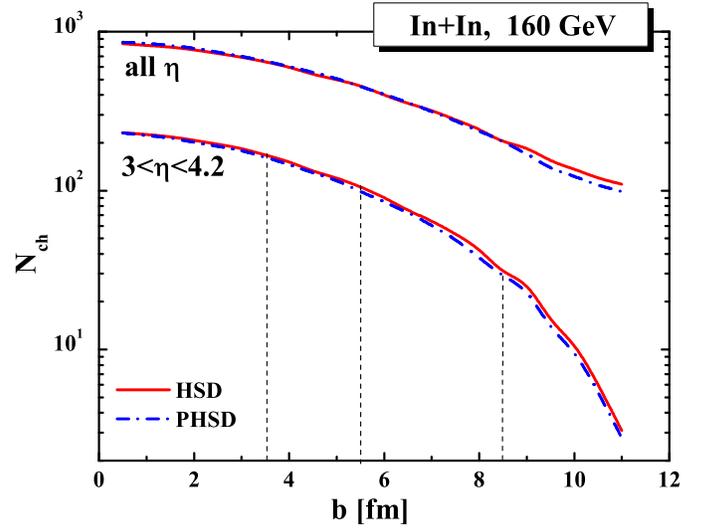}
 \caption{(color online) Number of charged particles as a
function of the impact parameter from HSD (solid lines) and PHSD
(dash-dot lines) integrated over rapidity (upper lines) and within
the pseudo-rapidity acceptance window of the NA60 experiment (lower
lines). The vertical dashed lines indicate the different centrality
classes. } \label{Nch}
\end{figure}

Next we investigate the centrality dependance of the dilepton
production in heavy-ion collisions as SPS energies. In
Fig.~\ref{centrality} we present the mass spectra of excess dimuons
from $In+In$ at 158~AGeV for $0.2<p_T<2.4$~GeV and $3<\eta<4.2$ from
PHSD for different centrality bins. The dash-dotted, dashed, and
solid lines show, respectively, the dilepton yield from the
in-medium $\rho$ with a broadened spectral function, the dilepton
yield from the $q+\bar q$ annihilation and the sum of them. We have
chosen the following centrality classes: central collisions (impact
parameter $0.5$~fm$<b<3.5$~fm), semi-central ($3.5$~fm$<b<5.5$~fm),
and semi-peripheral ($5.5$~$<b<8.5$~fm). The predictions in
Fig.~\ref{centrality} can be verified/falsified in the future by a
direct comparison to the data as the latter become available.

The yields in Fig.~\ref{centrality} are normalized to the number of
charged particles $N_{ch}$. By studying the dependence of $N_{ch}$
on the centrality in $In+In$ collisions at 158~AGeV in
Fig.~\ref{Nch} we find that PHSD and HSD give very similar results
(with only 5\% quantitative difference). This finding is in line
with the conclusions of the extended study in Ref~\cite{CasBrat}
that the multiplicities, rapidity- and
transverse-momentum-distributions of the non-strange particles
produced in heavy-ion collisions at 158~A$\cdot$GeV are only weekly
sensitive to the presence of a partonic phase with a (cross over)
phase transition. The average numbers of charged particles per unit
of pseudo-rapidity in PHSD and HSD for the chosen centrality classes
are shown in the following table 1:
\begin{center}
\begin{tabular}{| c | c | c | }
  \hline
  & \multicolumn{2}{|c|}{$<dN_{ch}/d\eta>$} \\
  Centrality & \phantom{P}HSD & PHSD \\
  \hline
  $\phantom{0.5<}b<8.5$ fm & 83.44 & 79.00 \\
  $0.5<b<3.5$  fm & 166.6 & 157.1 \\
  $3.5<b<5.5$  fm & 119.5 & 112.6 \\
  $5.5<b<8.5$  fm & 58.13 & 55.54 \\
  \hline
\end{tabular}
\end{center}
%
{Table 1: \small The average numbers of charged particles per unit
of pseudo-rapidity in PHSD and HSD for the different centrality
classes. \vspace{12pt}}

\begin{figure*}
\includegraphics[width=0.97\textwidth]{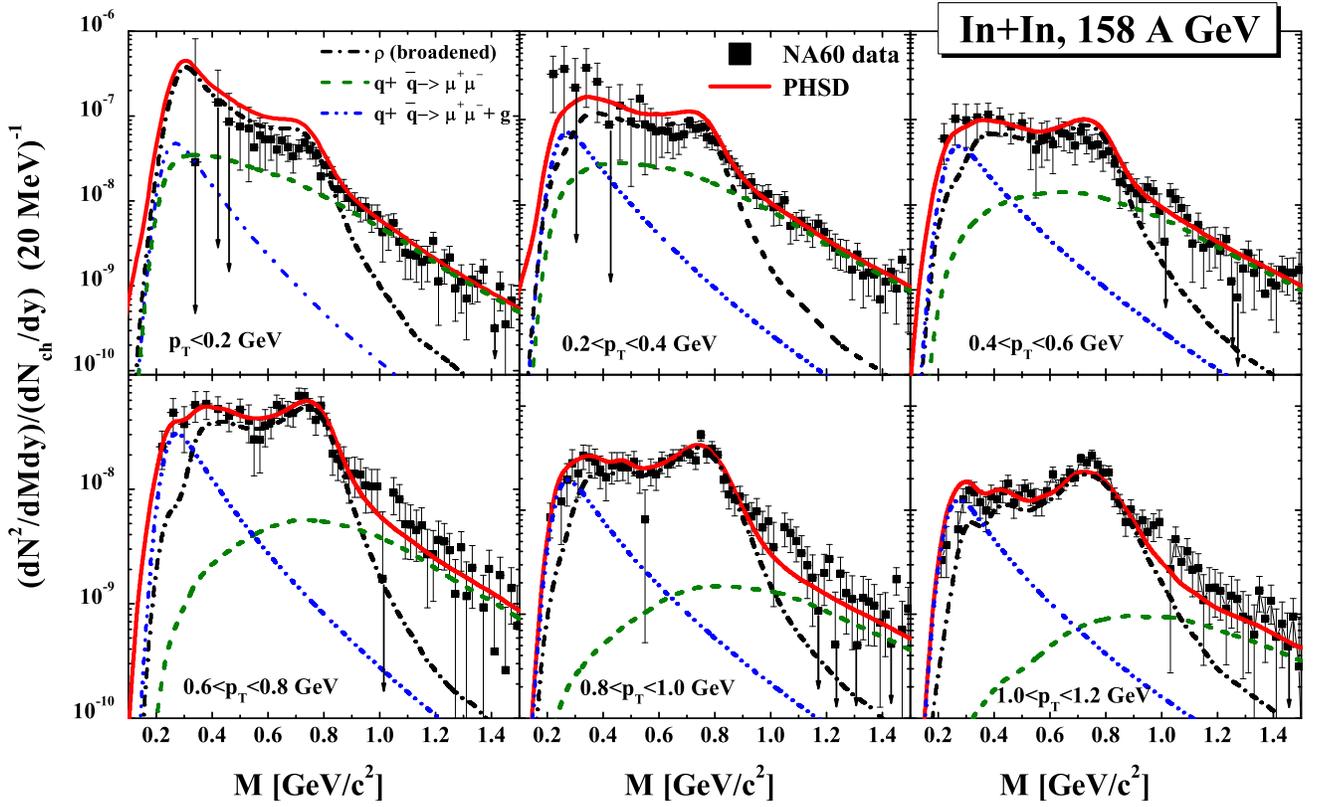}
\caption{(color online) Acceptance corrected mass spectra of excess
dimuons from $In+In$ collisions at 158~A$\cdot$GeV for different
transverse momentum bins from PHSD compared to the data of the NA60
Collaboration~\protect{\cite{NA60,Arnaldi:2008er}}. The dash-dotted,
dashed/dash-dot-dot, and solid lines show, respectively, the
dilepton yield from the in-medium $\rho$ with a broadened spectral
function, the dilepton yield from  $q+\bar q$ and the sum of them.}
\label{pTbins}
\end{figure*}

The NA60 Collaboration has accessed the information on the
transverse momentum dependence of dilepton production by measuring
the dilepton yield in different bins of $p_T$. In Fig.~\ref{pTbins}
we show the mass spectra of excess dimuons from $In+In$ at 158~AGeV
for different transverse momentum bins from PHSD compared to the
data of the NA60 Collaboration~\protect{\cite{NA60,Arnaldi:2008er}}.
The dash-dotted, dashed/dash-dot-dot, and solid lines show,
respectively, the dilepton yield from the in-medium $\rho$ with a
broadened spectral function, the dilepton yield from the $q+\bar q$
annihilation and the sum of them. One observes a generally good
agreement with the data.

\begin{figure} \vspace{-0.5cm}
\includegraphics[width=0.5\textwidth]{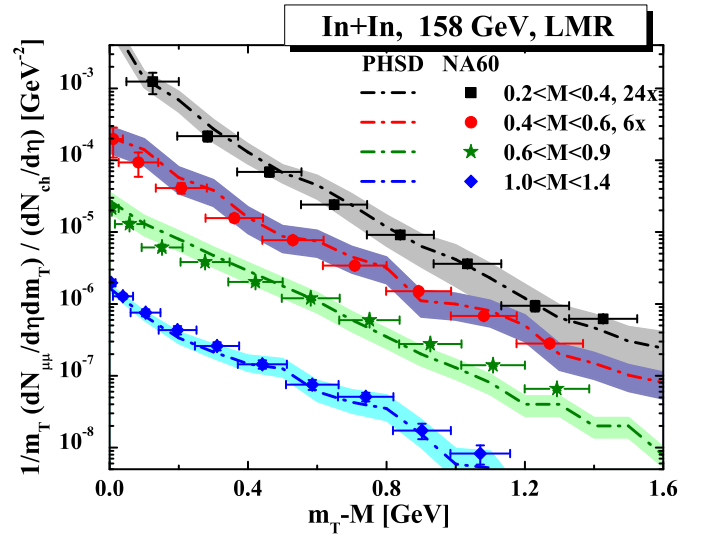}
\caption{(color online) Transverse mass spectra of dileptons for
In+In at 158 A$\cdot$GeV in PHSD compared to the data of the NA60
Collaboration~\protect{\cite{NA60,Arnaldi:2008er}}.} \label{Slopes1}
\end{figure}

\begin{figure}
\includegraphics[width=0.5\textwidth]{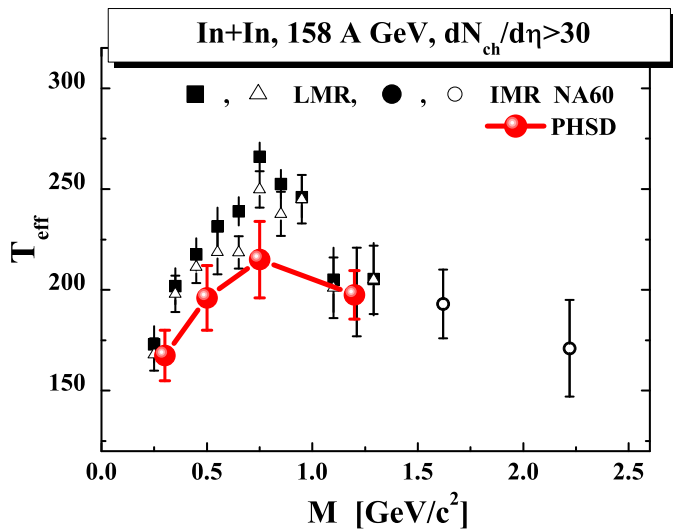}
\caption{(color online) The inverse slope parameter $T_{eff}$ of the
dimuon yield from In+In at 158 A$\cdot$GeV as a function of the
dimuon invariant mass in PHSD compared to the data of the NA60
Collaboration~\protect{\cite{NA60,Arnaldi:2008er}}.} \label{Slopes2}
\end{figure}

In Fig.~\ref{Slopes1}, transverse mass spectra of dileptons for
In+In at 158~AGeV in PHSD are compared to the data of the NA60
Collaboration for the four mass bins. The comparison of the mass
dependance of the slope parameter evolution in PHSD and the data is
shown explicitly in Fig.~\ref{Slopes2}.
Including partonic dilepton sources allows us to reproduce in PHSD
the $m_T$-spectra (cf. Fig.~\ref{Slopes1}) as well as the finding of
the NA60 Collaboration~\cite{NA60,Arnaldi:2008er} that the effective
temperature of the dileptons (slope parameters) in the intermediate
mass range is lower than that of the dileptons in the mass bin
$0.6<M<1$~GeV, which is dominated by hadronic sources (cf.
Fig.~\ref{Slopes2}).
The softening of the transverse mass spectrum with growing invariant
mass implies that the partonic channels occur dominantly before the
collective radial flow has developed. Also, the fact that the slope
in the lowest mass bin and the highest one are approximately equal
-- both in the data and in PHSD -- can be traced back to the two
windows of the mass spectrum that in our picture are influenced by
the radiation from the sQGP: $M=0-0.6$~GeV and $M>1$~GeV (cf. the
discussions of Fig.~\ref{Thermal} and Fig.~\ref{NA60_AC}).
A detailed look at the PHSD results shows that in total we still
slightly underestimate the slope parameter $T_{eff}$ in the
$\rho$-mass region which might be due to missing partonic initial
state effects or an underestimation of flow in the initial phase of
the reaction.


\section{Summary}

To address dilepton production in a hot and dense medium as created
in heavy-ion collisions, we have employed an up-to-date relativistic
transport model, the Parton-Hadron-String
Dynamics~\cite{CasBrat,BrCa11} (PHSD). PHSD consistently describes
the full evolution of a relativistic heavy-ion collision from the
initial hard scatterings and string formation through the dynamical
deconfinement phase transition to the quark-gluon plasma as well as
hadronization and to the subsequent interactions in the hadronic
phase.

In the present work, we have studied the dilepton production in
In+In collisions at 158~A$\cdot$GeV within the  PHSD off-shell
transport approach including a collisional broadening of vector
mesons, microscopic secondary multi-meson channels and the strongly
interacting QGP radiation, which is described by the interactions of
dynamical quasiparticles in line with the degrees of freedom
propagated in the transport approach.

A comparison to the data of the NA60 Collaboration shows that the
dilepton yield is well described by including the collisional
broadening of vector mesons, while simultaneously accounting for the
electromagnetic radiation of the strongly coupled quark-gluon plasma
(sQGP) via off-shell quark-antiquark annihilation, quark
annihilation with gluon Bremsstrahlung and the gluon-Compton
scattering mechanisms.

In particular, the spectra in the intermediate mass range
(1~GeV~$\le M\le2.5$ GeV) are found to be dominated by
quark-antiquark annihilation in the nonperturbative QGP. Also, the
observed softening of the transverse mass spectra at intermediate
masses (1~GeV~$\le M\le2.5$~GeV) is approximately reproduced.

Furthermore, for dileptons of low masses ($M<0.6$~GeV), a sizable
contribution of partonic processes (in particular, the quark
annihilation with the gluon bremsstrahlung) is found, thus possibly
providing another window for probing the properties of the sQGP.

Our present findings will have to be controlled by dilepton
measurements at RHIC and LHC energies, since the PHSD approach is
designed to operate also at these higher energies. Our
results/predictions for higher energies will be presented in near
future.

\section*{Acknowledgements}

O.L. and E.L.B. acknowledge financial support through the HICforFAIR
framework of the LOEWE program. V.O. acknowledges financial support
of the H-QM and HGS-Hire graduate schools. The work of C.M.K was
supported by the U.S. National Science Foundation under Grants No.
PHY-0758115 and No. PHY-1068572, the US Department of Energy under
Contract No. DE-FG02-10ER41682, and the Welch Foundation under Grant
No. A-1358, and he further would like to thank the Frankfurt
Institute for Advanced Studies for the warm hospitality during his
research visit.

\vspace{0cm}

\bibliographystyle{h-physrev3}
\bibliography{PHSDdilept}

\end{document}